\documentclass[letter]{aa}

\usepackage{graphicx}
\usepackage{txfonts}
\usepackage{hyperref}
\usepackage{multirow,bigdelim}
\usepackage{bm}

\newcommand{\bnabla}{\mbox{\boldmath$\nabla$}}
\newcommand{\bOmega}{\boldsymbol{\Omega}}
\newcommand{\iim}{{\rm i}}
\newcommand{\bu}{\boldsymbol{u}}
\newcommand{\bg}{\boldsymbol{g}}

\begin{document}

\title{Solar inertial modes:\\ Observations,  identification, and diagnostic promise}
\titlerunning{Solar inertial modes}

\author{Laurent~Gizon\inst{1,2,3} \and Robert~H.~Cameron\inst{1} \and Yuto~Bekki\inst{1} \and Aaron~C.~Birch\inst{1} \and Richard~S.~Bogart\inst{4} \and Allan~Sacha~Brun\inst{5} \and Cilia~Damiani\inst{1} \and Damien~Fournier\inst{1} \and Laura~Hyest\inst{1,6} \and Kiran~Jain\inst{7} \and B.~Lekshmi\inst{1} \and Zhi-Chao~Liang\inst{1} \and Bastian~Proxauf\inst{1} }

\offprints{L. Gizon, \email{gizon@mps.mpg.de}}

\institute{Max-Planck-Institut f\"ur Sonnensystemforschung,  37077~G\"ottingen, Germany 
    \and Institut f\"ur Astrophysik, Georg-August-Universit\"at G\"ottingen,  37077~G\"ottingen, Germany
            \and Center for Space Science, NYUAD Institute, New York University Abu Dhabi, Abu Dhabi, UAE
\and W.~W.~Hansen Experimental Physics Laboratory, Stanford University, Stanford CA~94305, USA
\and AIM, CEA, CNRS, Universités Paris et Paris-Saclay, 91191 Gif-sur-Yvette, Cedex, France
\and Institut Sup\'erieur de l'A\'eronautique et de l'Espace (ISAE-SUPAERO), 31400~Toulouse, France
\and National Solar Observatory, Boulder, CO 80303, USA}

\date{Submitted 1 June 2021; Revised 24 June 2021; Accepted 1 July 2021}

\abstract{
The oscillations of a slowly rotating star have long been classified into spheroi\-dal and toroidal modes. The spheroidal modes include the well-known 5-min acoustic modes used in helioseismology. Here we report observations of the Sun's toroidal modes, for which the restoring force is the Coriolis force and whose periods are on the order of the solar rotation period. By comparing the observations with the normal modes of a differentially rotating spherical shell, we are able to identify many of the observed modes. These are the high-latitude inertial modes, the critical-latitude inertial modes, and the equatorial Rossby modes. 
In the model, the high-latitude and critical-latitude modes have maximum kinetic energy density at the base of the convection zone, and the high-latitude modes are baroclinically unstable due to the latitudinal entropy gradient.
As a first application of inertial-mode helioseismology, we constrain the superadiabaticity and the turbulent viscosity in the deep convection zone.
}


\maketitle 

\noindent Movies and additional material can be downloaded from:\\
\noindent
\href{http://www2.mps.mpg.de/projects/seismo/SolarInertialModes/}{http://www2.mps.mpg.de/projects/seismo/SolarInertialModes/}.

\section{Introduction}

The free oscillations of a nonrotating spherical star  
have zero radial vorticity and are called spheroidal modes: they are the pressure (p), surface-gravity (f), and gravity (g) modes. The p~and f~modes, discovered on the Sun by \citet{Leighton1962}, are used to infer the structure and dynamics of the solar interior \citep{JCD2002}. The solar g~modes would also have important diagnostic potential regarding the radiative interior of the Sun; however, they evanesce in the convection zone and their amplitudes at the surface are exceedingly small \citep{Garcia2007, Alvan2015}.

When slow uniform rotation is included in the model, additional modes of oscillation become possible. In particular, quasi-toroidal modes that resemble classical Rossby modes, known as r~modes, are predicted \citep{Papaloizou1978}.
They owe their existence to the Coriolis force, have frequencies on the order of the rotation frequency, and propagate in the retrograde direction.
Adding the Sun's differential rotation introduces  critical latitudes where the phase speed of a mode is equal to the local rotation velocity. 
In the inviscid case, the eigenvalue problem is singular at the critical latitudes \citep{Watson1981, Charbonneau1999}.
Adding viscosity changes the eigenvalue problem from  second order to fourth order \citep[e.g.,][]{Baruteau2013JFM}. The singularity disappears and new quasi-toroidal modes appear, which are analogous to those of the plane Poiseuille viscous flow in classical hydrodynamics \citep[][and references therein]{Gizon2020-beta}. In the following, we loosely refer to the modes with frequencies on the order of the rotational frequency as inertial modes.

\begin{figure*}[!t] 
\centering
\includegraphics[width=0.69\linewidth]{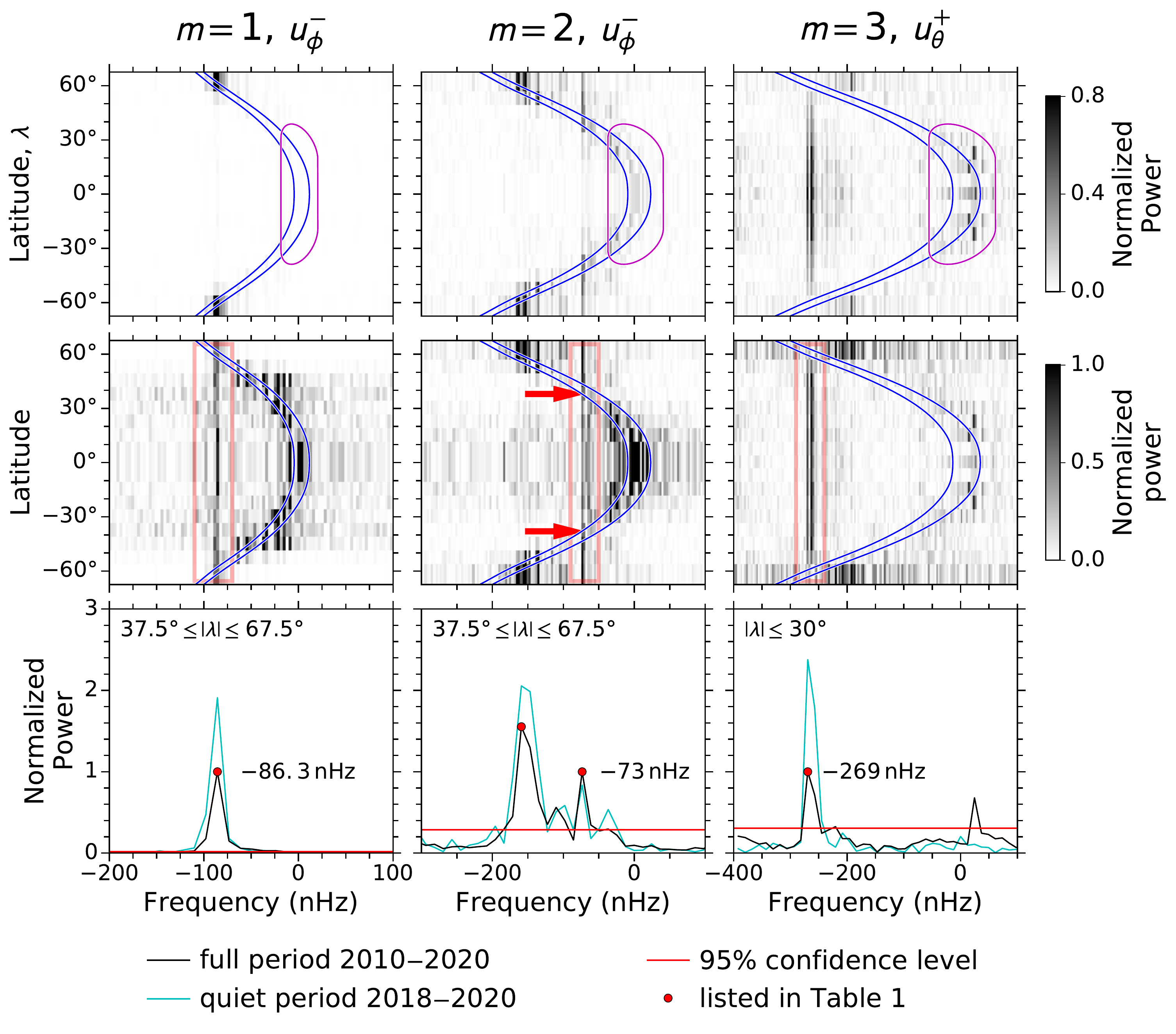}
\caption{
\label{fig.observations}
Power spectra showing selected modes of oscillation in the Carrington frame. 
Each column corresponds to a particular  $m$ and velocity component, as indicated at the top.
Each row shows a different representation of the power spectrum.  
In the top row, the power spectral density is plotted as a function of frequency and latitude. The two blue curves show $m(\Omega - \Omega_{\rm Carr})/2\pi$ at the surface and at $r = 0.95 R_\odot$, where $\Omega(r,\theta)$ is the solar angular velocity in the inertial frame. The purple contour delineates the region in frequency--latitude space affected by inflows into active regions, $m (\Omega_{\rm AR}-\Omega_{\rm Carr}) /2\pi$ (see Fig.~\ref{fig.ar_freq_range}). In the second row, the power at each latitude is normalized by its average value over the frequency range between the red bars; this shows that each mode has excess power over a large range of latitudes.
{The red arrows point to the critical latitudes of $\pm 38^\circ$ at the surface for the mode with frequency  $-73$~nHz.}
In the third row, the power is averaged over the selected latitude bands specified on the plots, and the frequency resolution is reduced to $12.24$ nHz. The red dots point to modes that are not activity-related (see text) and they are listed in Table~\ref{full-table}.
}
\end{figure*}

\begin{figure*}[!t]
\centering
\includegraphics[width=\linewidth]{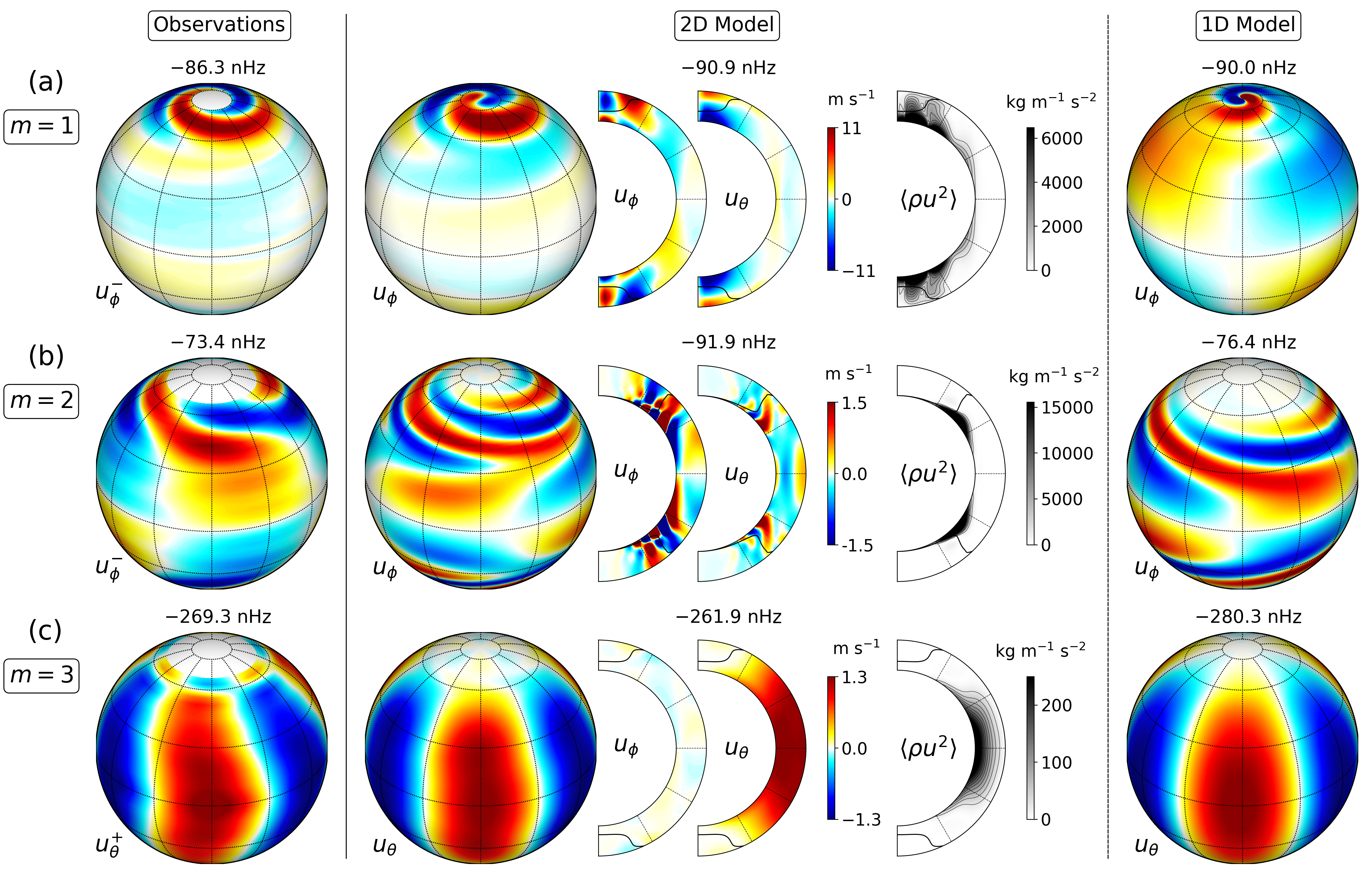}
\caption{
Observed and model eigenfunctions for the modes shown in Fig.~\ref{fig.observations}. 
{The left column shows the observed velocity ($u_\phi^-$ for the $m=1$ and $m=2$ modes, $u_\theta^+$ for the $m=3$ mode). The middle columns show the corresponding  eigenfunctions of the 2D~model for $\nu_{\rm t} = 100$~km$^{2}$~s$^{-1}$ and $\delta =0$},  at the surface and through the central meridian, together with the kinetic energy density. The thick black curves show the critical latitudes. The rightmost column shows the  eigenfunctions of the 1D~model at the surface.
The retrograde propagation of these modes in the Carrington frame is illustrated as an   
%
\href{https://edmond.mpdl.mpg.de/imeji/collection/oKXKmIl3VbdVYFok/item/J9lFY2lKtdloLQX}{\it online movie}.
The other velocity components are shown in Fig.~\ref{fig.eigenfunctions_complementary} and the radial vorticity is shown in Fig.~\ref{fig.vorticity}.
\label{fig.eigenfunctions}}
\end{figure*}

Inertial modes were detected  on some rapidly rotating stars \citep[see the review by][]{Aerts21}. The search for the Sun's inertial modes requires observations over many times the $27$-day solar rotation period due to their low frequencies and amplitudes. 
Equatorial Rossby modes modified by the solar differential rotation have already been reported \citep{Loeptien2018}.
Here we report observations of a rich spectrum of inertial modes of the Sun over a wide range of latitudes, and we show they can be used to directly probe the superadiabaticity and turbulent viscosity in the deep convection zone. 
The degree to which the lower half of the convection zone 
is superadiabatic (or subadiabatic) is important in the context of storing the toroidal magnetic field so that it can build up over the course of the $11$-year solar cycle \citep{Hotta2017}. 
The turbulent viscosity is one of the important turbulent transport processes that acts in combination with the observed meridional flow \citep{Gizon-MC-2020} to explain the equatorward drift of the latitudes at which sunspots emerge \citep{Cameron2016}. 

By definition, a normal mode is separable in time and space; it is characterized by a single eigenfrequency that is independent of position and by a displacement eigenfunction that is independent of time. Working in the frequency--latitude domain is key to the observational discovery and the identification of the quasi-toroidal normal modes of the Sun.
\section{Observations}

 We use helioseismic maps of horizontal flows
 near the solar surface provided by the Stanford ring-diagram pipeline \citep{Bogart2011a, Bogart2011b} applied to continuous high-resolution observations from the Helioseismic and Magnetic Imager (HMI) onboard the Solar Dynamics Observatory (SDO)  for the period from 1~May 2010 to 6~September 2020. The two horizontal flow components are standard data
products:  $u_\theta(\theta, \phi, t)$ in the colatitudinal direction and $u_\phi(\theta, \phi, t)$ in the longitudinal direction \citep[$\theta$ and $\phi$ increase southward and prograde, respectively; see][]{Proxauf2020}. 
The flows are measured either with a cadence of $d t = 27.28$~hr and an effective spatial resolution of $15^\circ$ in both coordinates, or with a more rapid cadence of $d t/3$ and a finer spatial resolution of $5^\circ$; the spatial sampling  is half the resolution such that there is a $50$\% overlap between neighboring measurements.
The highest latitude is $67.5^\circ$ for the low-resolution maps and $80.0^\circ$ for the high-resolution maps. The longitude, $\phi$, is defined in the Carrington frame of reference, which rotates at the frequency $\Omega_{\rm Carr}/2\pi = 456.0$~nHz with respect to an inertial frame (close to the equatorial rotation rate at the surface). The zero and yearly frequencies were removed from the data.

The structure of the Sun and its differential rotation is nearly symmetric with respect to the solar equator. Consequently, modes with a toroidal component can be called either symmetric or antisymmetric depending on the north--south symmetry of the surface radial vorticity. This terminology has been used before in the literature \citep{Charbonneau1999}. A symmetric mode has a symmetric $u_\theta$ and antisymmetric $u_\phi$, while an antisymmetric mode has an antisymmetric $u_\theta$ and symmetric $u_\phi$.
After symmetrizing (superscript "$+$") or anti-symmetrizing ("$-$") the data with respect to the equator, we computed the Fourier transform of the two velocity components in longitude and in time,
\begin{equation}
\hat{u}^\pm_j(\theta, m, \omega) = 
\sum_{\phi, t} 
u^\pm_j(\theta, \phi, t)\ {\rm e}^{-\iim( m\phi-\omega t )}    ,
\end{equation}
where $j$ is either $\theta$ or $\phi$,  $\omega$ is the angular frequency, $m$ is the integer longitudinal wavenumber, and the sums were taken over all longitudes and all times.  We considered frequencies in the range $|\omega / 2 \pi| \leq 400$~nHz and $m$ in the range from $1$ to $10$ to focus on the large-scale motions. 
For each choice of velocity component~$j$, symmetry~$s$,  and wavenumber $m$, the power spectral density $PSD =  |\hat{u}_j^s(\theta, m, \omega)|^2$ is a function of colatitude and frequency.

For illustration purposes, we show the detection of three  global modes of oscillation in the inertial frequency range in Fig.~\ref{fig.observations} ($15^\circ$ resolution). 
For each mode, there is clear excess power at the same frequency over a range of latitudes. Several  types of modes can be seen. 
The symmetric $m = 1$
mode at a frequency near $-86$~nHz is visible at all latitudes in the power spectrum; it
has most of its power at latitudes of $50^\circ$ and above (the $5^\circ$ observations show that the power keeps increasing with latitude up to at least $80^\circ$).
It corresponds to the high-latitude velocity features previously reported \citep{Hathaway2013, Bogart2015, Hathaway2021}, although it was not recognized as a normal mode of the whole convection zone. 
The second example is the symmetric $m = 2$ mode of oscillation at $-73$~nHz. This mode is also seen over the entire latitude range of the observations, but it has most of its power  concentrated near the critical latitude of $38^\circ$ (see Fig.~\ref{fig.observations}). The power is strong above the critical latitude, but decreases toward the poles.
The third example is  the $m = 3$ equatorial Rossby modes \citep{Loeptien2018} at a frequency of $-269$~nHz, for which the power is mostly confined to lie between the critical latitudes ($\pm 59^\circ$ for this mode's frequency) where the mode is trapped \citep{Gizon2020-beta}.

We have detected many tens of normal modes of oscillation at low frequencies, as shown in Figs.~\ref{fig.overview.1}\,--\,\ref{fig.overview.10} and reported in Table~\ref{full-table}. 
These modes are associated with significant (above $95\%$ confidence level)  excess power in at least one of three latitude bands (low latitudes below $30^\circ$, mid latitudes from $15^\circ$ to $45^\circ$, and high latitudes from $37.5^\circ$ to $67.5^\circ$). 
While the most striking features in the power spectra are narrow peaks, a closer inspection reveals ranges in frequency and latitude of additional excess power. 
For example, the $m = 8$ power spectrum for $u_\theta^+$ (Fig.~\ref{fig.m8}) has excess power at low latitudes at frequencies between $-135$~nHz and $-65$~nHz, which can be attributed to the presence of a dense spectrum of modes adjacent to the equatorial Rossby mode. 

To avoid misidentifying active-region inflows \citep{Gizon2001} as modes of oscillation in the power spectra, we defined a region in frequency--latitude space based on the active-region rotation rates and latitudes  \citep{Kutsenko2021}, as shown in Fig.~\ref{fig.ar_freq_range}. 
A peak in the power spectrum for the entire observation period is not reported in Table~\ref{full-table} if it is in the activity area and does not have significant power during the quiet-Sun period (February~2018 to September~2020), see Figs.~\ref{fig.overview.1}\,--\,\ref{fig.overview.10}. 
We also checked that the reported modes were not misidentified due to leakage from the window function \citep{Liang2019}, and that they are also seen in the Global Oscillation Network Group (GONG) data (Fig.~\ref{fig.gong}).

For each mode, we extracted the two velocity components of a mode eigenfunction in a narrow frequency range around the mode frequency within one linewidth \citep{Proxauf2020}. Examples of the surface eigenfunctions are shown in the left column of Fig.~\ref{fig.eigenfunctions}. 
 
\section{Mode identification}

To identify the observed modes 
of oscillation, we computed the eigenmodes of a spherical shell with $0.710 \leq r/R_\odot \leq 0.985$ that rotates like the Sun. The internal rotation rate is specified by p-mode helioseismology \citep{Larson2018}. For the sake of simplicity, we chose the model to have very few free parameters: a constant fluid viscosity $\nu_{\rm t}$
and a constant superadiabaticity  $\delta$, which gives the degree of convective instability.
Each mode eigenfunction is  proportional to  $\exp(\iim m \phi - \iim \sigma t)$, where $\sigma$ is the complex eigenfrequency. 
For each $m$, we solved the 2D (\textit{r}--$\theta$)
eigenvalue problem (Appendix~\ref{app.2dsolver}). 
In order to highlight the main physics, we also computed the purely toroidal (horizontal)  modes at fixed radius $r=R_\odot$. For each $m$, we solved the 1D ($\theta$) eigenvalue problem where the only free parameter is the turbulent viscosity (Appendix~\ref{app.1dsolver}). The oscillation spectrum of this 1D model is much less cluttered (no radial overtones for the inertial modes and no convective modes).

\begin{figure}
\centering
\includegraphics[width=\linewidth]{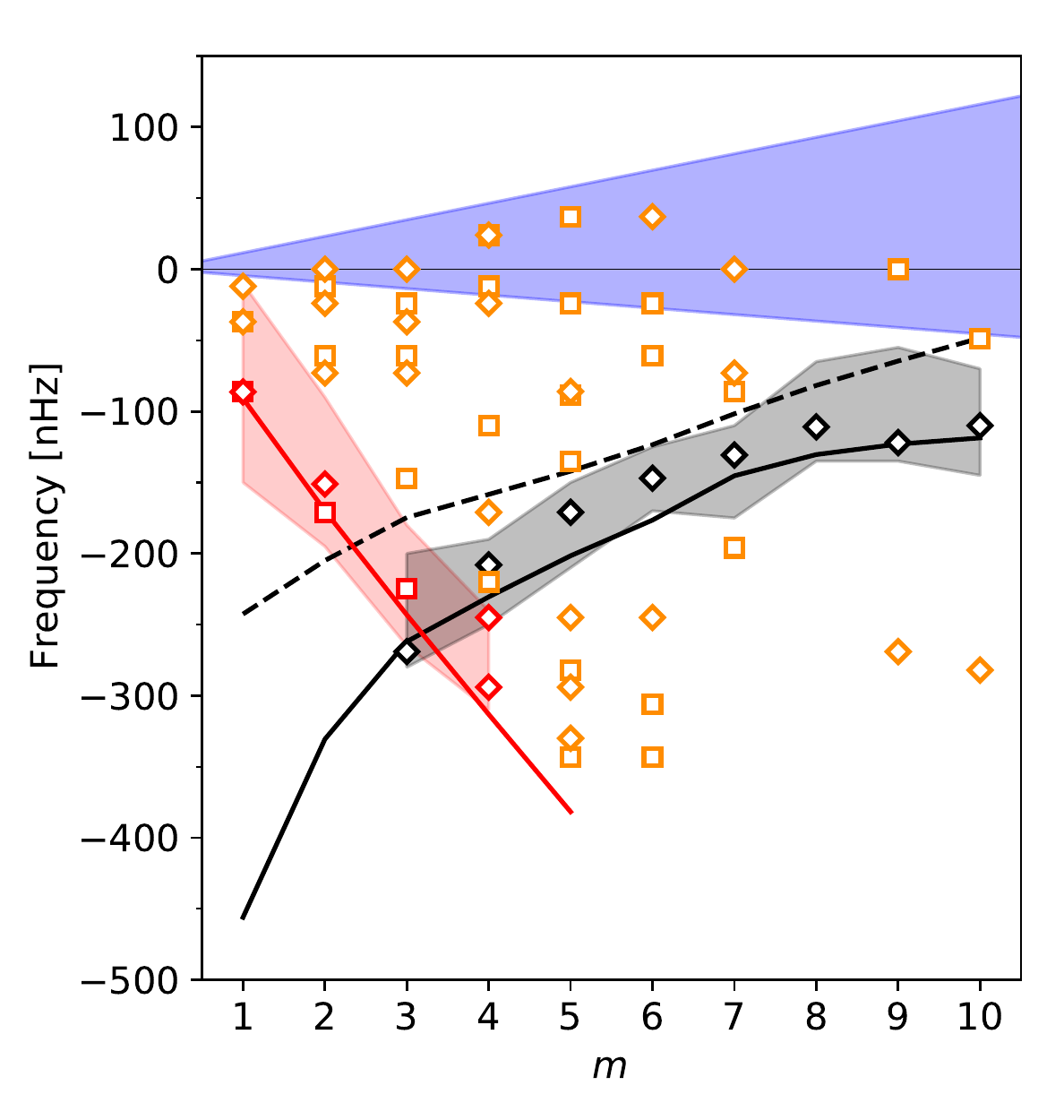}
\caption{Mode frequencies in the Carrington frame for the observations and the 2D model {(Re\ $\sigma$)}. The symbols show the observed modes (diamonds for symmetric modes and squares for antisymmetric modes). The red symbols show the high-latitude modes, the orange symbols the critical-latitude modes, and the black symbols the equatorial Rossby modes. 
The rose- and gray-shaded areas show the observed frequency ranges of excess power (last column of Table~\ref{full-table}). 
For reference, the blue-shaded area gives the range of rotation rates at the equator between the surface and $0.95 R_\odot$. 
The curves give the dispersion relations for the modes of the 2D model with {$\nu_{\rm t} = 100$~km$^2$~s$^{-1}$ and $\delta=0$}. The red curve is the dispersion relation for the high-latitude modes. The solid and dashed-black curves are for the fundamental ($n = 0$) and first overtone ($n = 1$) equatorial Rossby modes. 
\label{fig.dispersion} 
}
\end{figure}

We then sought a match with the observed modes. We did not tune the parameters of the 2D model to match the observations exactly; we performed a sensitivity study using $\delta
= -10^{-6}$, $-2\times 10^{-7}$, $0$, $2\times 10^{-7}$, $10^{-6}$ (Fig.~\ref{fig.changedelta}) and $\nu_{\rm t}$ = 50, 100, 250, 500 km$^2$~s$^{-1}$ (Fig.~\ref{fig.changenu}). 
We found that $\delta = 0$ and $\nu_{\rm t} = 100$~km$^2$~s$^{-1}$ provide a good match (Fig.~\ref{fig.eigenfunctions}) for the surface eigenfunctions and eigenfrequencies of the three modes of Fig.~\ref{fig.observations}. As part of the identification, we sought modes of the models that have long lifetimes or are growing (Fig.~\ref{fig.eigenvalues-m1}). The identified modes are representatives of three main families of modes: the high-latitude inertial modes (Fig.~\ref{fig.eigenfunctions}a), the critical-latitude inertial modes (Fig.~\ref{fig.eigenfunctions}b), and the equatorial Rossby modes (Fig.~\ref{fig.eigenfunctions}c). This classification is supported by the dispersion relations at small $m$ (Fig.~\ref{fig.dispersion}).

The high-latitude inertial modes are analogous to the "wall modes" in  plane Poiseuille flows \citep{Gizon2020-beta}. They are seen in both the 1D and 2D eigenvalue problems for $m \leq 5$. In the 2D model, the eigenfunctions are dominantly toroidal and extend to the bottom of the convection zone, with their highest kinetic energy density near the base of the convection zone (Fig.~\ref{fig.eigenfunctions}a).
This is unlike the kinetic energy density of the p modes, which always peaks near the surface. The correct tilt of the spiral structure is only obtained in the 2D model (Fig.~\ref{fig.eigenfunctions}a). In this model, the high-latitude modes become baroclinically unstable (Fig.~\ref{fig.eigenvalues-m1}) due to the latitudinal entropy gradient resulting from the thermal wind balance
 (\citeauthor{Knobloch1982}  \citeyear{Knobloch1982}; Bekki et al., in prep.).\footnote{The formation of a spiral at high latitudes  by baroclinic instability has also been discussed in the context of Venus' atmosphere \citep{Kashimura2019}.}
In the 1D model, only the $m=1$ high-latitude modes are self-excited (Fig.~\ref{fig.eigenvalues-m1}) as a result of a shear instability at high latitudes; however, the tilt of the spiral is not consistent with the observations.

Critical-latitude inertial modes are found 
for both the 1D and 2D models.
Their amplitudes are maximum near their critical latitudes; they are known as "center modes" in 1D hydrodynamics \citep{Gizon2020-beta}. The kinetic energy density of the $m = 2$ mode of the 2D model at $-92$ nHz ($-73$~nHz observed) is concentrated near the base of the convection zone near $45^\circ$ latitude (Fig.~\ref{fig.eigenfunctions}b). This is a very important place in the Sun, as it is where the toroidal magnetic field generation should be strongest \citep{Spruit2011}.

Equatorial Rossby modes are the easiest to identify: their frequencies are close to the classical dispersion relation for uniform rotation, $\sigma=-2 \Omega/(m+1)$.
The 2D model supports modes with a different number of nodes in the radial direction, $n$. The frequencies of the observed equatorial Rossby modes span the range between the $n = 0$ and the $n = 1$ branches of the dispersion relation (Fig.~\ref{fig.dispersion}). For example, the $m = 3$ mode is identified as a fundamental mode ($n = 0$).

\section{Conclusion}

We observed 
and identified three families of global-scale inertial modes in the solar convection zone, within the search range $|\omega/2\pi|\leq 400$~nHz and $1\leq m \leq 10$.
Some of these modes  are self-excited in the models. We also found extended regions in frequency space where closely packed modes exist. The  modes we have identified are sensitive to the physical conditions deep in the convection zone (see plots of the kinetic energy density in Fig.~\ref{fig.eigenfunctions}). The eigenfrequencies and surface eigenfunctions of the high- and critical-latitude inertial modes have diagnostic potential for the latitudinal entropy gradient, the superadiabaticity (Fig.~\ref{fig.changedelta}), and the turbulent viscosity (Fig.~\ref{fig.changenu}), which are largely unconstrained by traditional p-mode helioseismology. We find that the observed inertial modes are compatible with $\delta < 2\times 10^{-7}$ and $\nu_{\rm t} \le 100$~km$^2$~s$^{-1}$ at the bottom of the convection zone.
These observational upper limits  are substantially below the expectation from mixing length theory --- by approximately one order of magnitude each
\citep{JCD1996,Munoz2011} --- and they imply that the convective motions in the lower half of the convection zone are weak. 
This might correspond to the slightly subadiabatic conditions seen below $0.8\ R_\odot$ in recent numerical simulations of solar convection \citep[see e.g.][]{Hotta2017, Kaepylae2017, Bekki2017}. While our upper limit on the turbulent velocities ($\approx \sqrt{3 \nu_{\rm t} / \tau} \leq 11$~m~s$^{-1}$ for a correlation time $\tau = 1$~month) is well below the mixing length value, it is just above the lower limit required to drive solar differential rotation \citep[8~m~s$^{-1}$ according to][]{Miesch2012}.
A lower convection zone that is only marginally unstable (or even stable) would allow a flux transport dynamo to wind up and transport the magnetic field in this region.
We expect that the characteristics of the observed inertial modes, including amplitudes and lifetimes, will allow us to infer $\delta(r)$ and $\nu_{\rm t}(r)$ and understand in which regime of rotating convection the Sun operates \citep{Hindman2020}.

\begin{acknowledgements}
Author contributions: This project was initiated and supervised by LG and RHC. The helioseismic ring parameter fits were provided by RSB for HMI and KJ for GONG. BP and Z-CL measured the mode parameters. YB solved the 2D~eigenvalue problem. DF and LH solved the 1D~eigenvalue problem. LG, RHC and ACB wrote the draft paper. All authors contributed to the final manuscript. 
We are very grateful to John~Leibacher for useful comments. {LG thanks the Max Planck Institute for Astrophysics for the opportunity to present a preliminary account of these results at the 2020 Biermann Lectures.} The HMI data are courtesy of NASA/SDO and the HMI Science Team. This work utilizes GONG data from the National Solar Observatory (NSO), which is operated by AURA under a cooperative agreement with NSF and with additional financial support from NOAA, NASA, and USAF. This work was supported in part by NASA contract NAS5-02139 to Stanford University. YB is a member of the International Max Planck Research School for Solar System Science at the University of G\"ottingen, and acknowledges partial support from the Japan Student Services Organization (JASSO). We acknowledge support from ERC Synergy Grant WHOLE~SUN 810218. LG acknowledges support from NYUAD Institute Grant G1502. LG, DF and BP acknowledge funding by  Deutsche Forschungsgemeinschaft (DFG, German Research Foundation) through SFB~1456/432680300 Mathematics of Experiment, project C04. The computational resources were provided by the German Data Center for SDO through German Aerospace Center (DLR) grant 50OL1701. LG, ACB, and CD acknowledge support from DLR under PLATO Data Center grant 50OO1501. 
\end{acknowledgements}

\bibliographystyle{aa}
\bibliography{biblio}

\onecolumn

\begin{appendix}

\section{Parameters of observed modes}
\begin{table*}[th!]
\caption{\label{full-table}
Solar inertial modes detected in HMI ring-diagram flow maps for 2010--2020. Frequencies are defined in the Carrington frame.  
}
\centering \tiny 
\begin{tabular}{c c c c c c c c c@{}l@{}l}

\hline\hline
$m$  & $(j,s)$ & mode frequency \tablefootmark{a}   & significance \tablefootmark{b} & critical latitude \tablefootmark{c} & latitude at max($u^s_j$)   & max($u_\theta$) & max($u_\phi$) & linewidth & \multicolumn{2}{c}{excess power range} \\
    &         &  [nHz]     &              & at $r=R_\odot$    & [multiples of $7.5^\circ$] & [m s$^{-1}$]    & [m s$^{-1}$]  & [nHz]     & \multicolumn{2}{c}{(not activity related)}                   \\
\hline
\multicolumn{10}{l}{Equatorial Rossby modes:} \\
3 & $(\theta,+)$ & $-269$ \tablefootmark{d} & $> 8$ $\sigma_\mathrm{L}$ & $59^\circ$ & $0^\circ$ & 1.5 & 1.2 & $< 24$ & & [$-280,-200$]~nHz \\
4 & $(\theta,+)$ & $-208$ \tablefootmark{d} & $7.5$ $\sigma_\mathrm{L}$ & $45^\circ$ & $0^\circ$ & 1.6 & 1.1 & $< 37$ & & [$-250,-190$]~nHz \\
5 & $(\theta,+)$ & $-171$ \tablefootmark{d} & $> 8$ $\sigma_\mathrm{L}$ & $37^\circ$ & $0^\circ$ & 1.0 & --- & $< 12$ & & [$-210,-150$]~nHz \\
6 & $(\theta,+)$ & $-147$ \tablefootmark{d} & $5.7$ $\sigma_\mathrm{L}$ & $31^\circ$ & $0^\circ$ & 1.3 & --- & $< 24$ & & [$-170,-125$]~nHz \\
7 & $(\theta,+)$ & $-130.7\pm2.8$ \tablefootmark{d,e} & $> 8$ $\sigma_\mathrm{L}$ & $25^\circ$ & $7.5^\circ$ & 1.5 & --- & $9.9\pm2.4$ & & [$-175,-110$]~nHz \\
8 & $(\theta,+)$ & $-110.9\pm2.4$ \tablefootmark{d,e} & $> 8$ $\sigma_\mathrm{L}$ & $22^\circ$ & $0^\circ$ & 2.0 & --- & $10.6\pm1.0$ & & [$-135,-65$]~nHz \\
9 & $(\theta,+)$ & $-122$ \tablefootmark{d} & $> 8$ $\sigma_\mathrm{L}$ & $22^\circ$ & $0^\circ$ & 1.1 & --- & $< 12$ & & [$-135,-55$]~nHz \\
10 & $(\theta,+)$ & $-110$ \tablefootmark{d} & $> 8$ $\sigma_\mathrm{L}$ & $19^\circ$ & $0^\circ$ & 1.4 & --- & $< 24$ & & [$-145,-70$]~nHz \\
\hline
\multicolumn{10}{l}{High-latitude inertial modes:} \\
1 & $(\phi,+)$ & $-86$ & $> 8$ $\sigma_\mathrm{H}$ & $58^\circ$ & $\geq67.5^\circ$ & --- & 2.5 & $< 12$ & & [$-110,-50$]~nHz \\
1 & $(\phi,-)$ & $-86.3\pm1.6$ 
 \tablefootmark{e,f} & $> 8$ $\sigma_\mathrm{H}$ & $58^\circ$ & $\geq67.5^\circ$ & 3.0 & 9.8 & $7.8\pm0.2$ & & [$-150,-10$]~nHz \\
2 & $(\phi,+)$ & $-171$ \tablefootmark{f} & $> 8$ $\sigma_\mathrm{H}$ & $58^\circ$ & $60^\circ$ & --- & 2.3 & $< 12$ & & [$-195,-100$]~nHz \\
2 & $(\phi,-)$ & $-151.1\pm4.3$ \tablefootmark{e,f} & $> 8$ $\sigma_\mathrm{H}$ & $56^\circ$ & $\geq67.5^\circ$ & 2.5 & 3.4 & $30.6\pm3.3$ & & [$-185,-90$]~nHz \\
3 & $(\phi,+)$ & $-224.7\pm2.5$ \tablefootmark{e,f} & $> 8$ $\sigma_\mathrm{H}$ & $53^\circ$ & $60^\circ$ & 1.6 & 1.8 & $9.7\pm1.7$ & & [$-265,-180$]~nHz \\
4 & $(\theta,+)$ & $-294$ & $3.9$ $\sigma_\mathrm{H}$ & $53^\circ$ & $\geq67.5^\circ$ & 1.0 & 0.8 & $< 12$ & \rdelim\}{2}{*}[] & \multirow{2}{*}{[$-310,-240$]~nHz} \\
4 & $(\theta,+)$ & $-245$ & $5.3$ $\sigma_\mathrm{H}$ & $49^\circ$ & $60^\circ$ & 1.1 & 1.1 & $< 24$ & & \\
5 & $(\theta,-)$ & $-343$ & $5.3$ $\sigma_\mathrm{H}$ & $52^\circ$ & $60^\circ$ & 0.7 & --- & $< 12$ & \rdelim\}{2}{*}[] & \multirow{2}{*}{[$-355,-275$]~nHz} \\
5 & $(\phi,+)$ & $-282$ & $2.6$ $\sigma_\mathrm{H}$ & $47^\circ$ & $52.5^\circ$ & 0.8 & 0.8 & $< 24$ & & \\
\hline
\multicolumn{10}{l}{Critical-latitude inertial modes:} \\
1 & $(\phi,+)$ & $-37$ & $> 8$ $\sigma_\mathrm{M}$ & $38^\circ$ & $37.5^\circ$ & --- & 1.3 & $< 24$ {\tablefootmark{g}} & & \\
1 & $(\phi,-)$ & $-37$ & $7.1$ $\sigma_\mathrm{M}$ & $38^\circ$ & $37.5^\circ$ & 0.5 & 0.9 & $< 12$ {\tablefootmark{g}} & & \\
1 & $(\phi,-)$ & $-12$ & $> 8$ $\sigma_\mathrm{L}$ & $20^\circ$ & $30^\circ$ & --- & 1.2 & $< 24$ & & \\
2 & $(\phi,+)$ & $-61$ & $6.8$ $\sigma_\mathrm{M}$ & $34^\circ$ & $52.5^\circ$ & --- & 1.1 & $< 24$ & \rdelim\}{2}{*}[] & \multirow{2}{*}{[$-65,0$]~nHz} \\
2 & $(\phi,+)$ & $-12$ & $> 8$ $\sigma_\mathrm{L}$ & $10^\circ$ & $22.5^\circ$ & 0.9 & 1.1 & $< 12$ & & \\
2 & $(\phi,-)$ & $-73$ & $> 8$ $\sigma_\mathrm{H}$ & $38^\circ$ & $45^\circ$ & 0.8 & 1.3 & $< 12$ & \rdelim\}{3}{*}[] & \multirow{3}{*}{[$-90,30$]~nHz} \\
2 & $(\phi,-)$ & $-24$ & $> 8$ $\sigma_\mathrm{M}$ & $20^\circ$ & $22.5^\circ$ & 0.9 & 1.4 & $< 24$ & & \\
2 & $(\phi,-)$ & $\phantom{-}0$ & $7.2$ $\sigma_\mathrm{L}$ & n/a & $7.5^\circ$ & --- & 1.0 & $< 12$ {\tablefootmark{g}} & & \\
3 & $(\phi,+)$ & $-147$ & $4.0$ $\sigma_\mathrm{H}$ & $44^\circ$ & $45^\circ$ & --- & 0.9 & $< 12$ & & \\
3 & $(\theta,-)$ & $-61$ & $3.6$ $\sigma_\mathrm{M}$ & $28^\circ$ & $37.5^\circ$ & 0.7 & 0.8 & $< 24$ & & \\
3 & $(\phi,+)$ & $-24$ & $> 8$ $\sigma_\mathrm{M}$ & $15^\circ$ & $15^\circ$ & --- & 1.0 & $< 12$ & & [$-50,10$]~nHz \\
3 & $(\phi,-)$ & $-73$ \tablefootmark{h} & $3.0$ $\sigma_\mathrm{M}$ & $31^\circ$ & $30^\circ$ & --- & 0.7 & $< 12$ & & \\
3 & $(\phi,-)$ & $-37$ & $6.6$ $\sigma_\mathrm{L}$ & $20^\circ$ & $22.5^\circ$ & --- & 1.0 & $< 24$ & \rdelim\}{2}{*}[] & \multirow{2}{*}{[$-50,30$]~nHz} \\
3 & $(\phi,-)$ & $\phantom{-}0$ & $6.6$ $\sigma_\mathrm{L}$ & n/a & $15^\circ$ & --- & 1.0 & $< 24$ {\tablefootmark{g}} & & \\
4 & $(\phi,+)$ & $-220$ & $4.3$ $\sigma_\mathrm{H}$ & $46^\circ$ & $45^\circ$ & --- & 0.5 & $< 12$ & & \\
4 & $(\phi,+)$ & $-110$ & $4.4$ $\sigma_\mathrm{M}$ & $33^\circ$ & $37.5^\circ$ & --- & 0.6 & $< 12$ & \rdelim\}{3}{*}[] & \multirow{3}{*}{[$-120,35$]~nHz} \\
4 & $(\phi,+)$ & $-12$ & $4.3$ $\sigma_\mathrm{L}$ & n/a & $15^\circ$ & 0.6 & 1.0 & $< 12$ {\tablefootmark{g}} & & \\
4 & $(\phi,+)$ & $\phantom{-}24$ & $5.5$ $\sigma_\mathrm{L}$ & n/a & $0^\circ$ & --- & 1.4 & $< 24$ & & \\
4 & $(\phi,-)$ & $-171$ & $3.4$ $\sigma_\mathrm{H}$ & $41^\circ$ & $45^\circ$ & --- & 0.6 & $< 12$ & & \\
4 & $(\phi,-)$ & $-24$ & $> 8$ $\sigma_\mathrm{L}$ & $10^\circ$ & $22.5^\circ$ & --- & 1.0 & $< 24$ & \rdelim\}{2}{*}[] & \multirow{2}{*}{[$-50,30$]~nHz} \\
4 & $(\phi,-)$ & $\phantom{-}24$ & $5.9$ $\sigma_\mathrm{L}$ & n/a & $7.5^\circ$ & --- & 1.0 & $< 24$ & & \\
5 & $(\phi,+)$ & $-135$  \tablefootmark{h} & $3.7$ $\sigma_\mathrm{M}$ & $32^\circ$ & $37.5^\circ$ & --- & 0.8 & $< 24$ & & \\
5 & $(\phi,+)$ & $-24$ & $3.5$ $\sigma_\mathrm{L}$ & $5^\circ$ & $15^\circ$ & --- & 1.0 & $< 24$ {\tablefootmark{g}} & & \\
5 & $(\theta,-)$ & $\phantom{-}37$ & $4.3$ $\sigma_\mathrm{L}$ & n/a & $15^\circ$ & 0.7 & --- & $< 24$ & & \\
5 & $(\theta,+)$ & $-330$ & $2.3$ $\sigma_\mathrm{H}$ & $51^\circ$ & $60^\circ$ & 0.7 & --- & $< 24$ & \rdelim\}{3}{*}[] & \multirow{3}{*}{[$-330,-190$]~nHz} \\
5 & $(\phi,-)$ & $-294$ & $2.1$ $\sigma_\mathrm{H}$ & $48^\circ$ & $52.5^\circ$ & --- & 0.7 & $< 24$ & & \\
5 & $(\phi,-)$ & $-245$ & $5.7$ $\sigma_\mathrm{H}$ & $44^\circ$ & $45^\circ$ & --- & 0.8 & $< 12$ & & \\
5 & $(\theta,+)$ & $-86$ & $4.0$ $\sigma_\mathrm{M}$ & $25^\circ$ & $37.5^\circ$ & 0.7 & --- & $< 12$ & & \\
6 & $(\theta,-)$ & $-343$  \tablefootmark{h} & $2.1$ $\sigma_\mathrm{H}$ & $47^\circ$ & $60^\circ$ & 0.5 & --- & $< 12$ & & \\
6 & $(\phi,+)$ & $-306$ & $3.3$ $\sigma_\mathrm{H}$ & $45^\circ$ & $60^\circ$ & --- & 0.4 & $< 12$ & & \\
6 & $(\theta,-)$ & $-61$ & $2.0$ $\sigma_\mathrm{M}$ & $18^\circ$ & $30^\circ$ & 0.6 & --- & $< 12$ & & \\
6 & $(\phi,+)$ & $-24$ & $4.5$ $\sigma_\mathrm{M}$ & n/a & $15^\circ$ & --- & 1.1 & $< 12$ & & \\
6 & $(\phi,-)$ & $-245$  \tablefootmark{h} & $2.7$ $\sigma_\mathrm{H}$ & $40^\circ$ & $45^\circ$ & --- & 0.5 & $< 12$ & & \\
6 & $(\phi,-)$ & $\phantom{-}37$ & $2.5$ $\sigma_\mathrm{L}$ & n/a & $7.5^\circ$ & --- & 0.7 & $< 12$ {\tablefootmark{g}} & & \\
7 & $(\theta,-)$ & $-196$ & $2.4$ $\sigma_\mathrm{H}$ & $33^\circ$ & $45^\circ$ & 0.7 & --- & $< 24$ \tablefootmark{g} & & \\
7 & $(\phi,+)$ & $-86$ & $3.6$ $\sigma_\mathrm{L}$ & $20^\circ$ & $22.5^\circ$ & --- & 0.6 & $< 12$ & & \\
7 & $(\theta,+)$ & $-73$ & $3.0$ $\sigma_\mathrm{M}$ & $18^\circ$ & $0^\circ$ & 0.5 & --- & $< 12$ & & \\
7 & $(\phi,-)$ & $\phantom{-}0$ & $3.6$ $\sigma_\mathrm{L}$ & n/a & $22.5^\circ$ & --- & 0.7 & $< 24$ & & \\
9 & $(\phi,+)$ & $\phantom{-}0$ & $5.3$ $\sigma_\mathrm{L}$ & n/a & $15^\circ$ & 0.7 & 0.6 & $< 12$ & & \\
9 & $(\theta,+)$ & $-269$ & $2.5$ $\sigma_\mathrm{H}$ & $34^\circ$ & $45^\circ$ & 0.5 & --- & $< 12$ & & \\
10 & $(\theta,-)$ & $-49$ & $3.3$ $\sigma_\mathrm{M}$ & $6^\circ$ & $22.5^\circ$ & 0.5 & --- & $< 12$ & & \\
10 & $(\theta,+)$ & $-282$ & $2.9$ $\sigma_\mathrm{H}$ & $33^\circ$ & $45^\circ$ & 0.3 & --- & $< 12$ & & \\
\hline
\end{tabular}
\tablefoot{
\tablefoottext{a}{The search range was limited to $1\leq m \leq 10$ and  $|\omega/2\pi |  \leq 400$  nHz.}  \tablefoottext{b}{The statistical significance of each peak is given in terms of the standard deviation computed for the most significant of the latitudinal averages shown in Figs.~\ref{fig.overview.1}--\ref{fig.overview.10} ($\sigma_L$ for low latitudes, $\sigma_M$ for mid latitudes, and $\sigma_H$ for high latitudes).}
\tablefoottext{c}{"n/a" means not applicable (no critical latitude at the surface, only deeper).}
\tablefoottext{d}{Mode reported by \citet{Loeptien2018}.}
\tablefoottext{e}{Mode parameters measured with a Lorentzian fit.}
\tablefoottext{f}{Frequency near that reported by \citet{Hathaway2021}.}
\tablefoottext{g}{Measured during the quiet Sun period 2018--2020.}
\tablefoottext{h}{Outside the activity frequency range, however not significant during the quiet Sun period 2018--2020.} 
}
\end{table*}

\section{Normal modes of the differentially rotating Sun}
\subsection{2D~eigenvalue solver}
\label{app.2dsolver}

In the Carrington frame, the linearized equations for the conservation of momentum, mass, and energy, together with the equation of state, are as follows:
\begin{align}
 &\rho D_t \bu' = - \bnabla p' + \rho' \bg - 2 \rho \mathbf{\Omega} \times \bu' - \rho r \sin\theta\ (\bu' \cdot \bnabla) \bOmega + \bnabla \cdot \bm{\mathcal{D}}, \label{eq.momentum}
\\
 & D_t \rho' = -\bnabla \cdot \left( \rho \bu'\right) , 
\\
 & D_t s' = \frac{c_p\ \delta}{H_p}u'_r - \frac{u'_{\theta}}{r}\frac{\partial s}{\partial\theta} + \frac{1}{\rho T} \bnabla \cdot \left( \kappa \rho T \bnabla s'\right) , 
\\ 
 & \frac{p'}{p} = \frac{\gamma\rho'}{\rho}+\frac{s'}{c_{v}} , 
\end{align}
where
\begin{equation}
D_t = \partial/\partial t + (\Omega - \Omega_{\rm Carr}) \partial/\partial \phi
\label{eq.materialDer}
\end{equation}
is the material derivative and $\Omega(r, \theta)$ is a differential rotation model close to the helioseismic measurements averaged over 2010\,--\,2020 \citep{Larson2018}. 
Linear perturbations are denoted with primes. 
The background model is based on a standard solar model \citep{JCD1996}, except for the superadiabaticity $\delta = \nabla - \nabla_{\rm ad}$, which is a constant parameter in the convection zone (the radiative zone is very stable below with $\delta \approx -0.1$). 
In the above equation, $H_p$ is the pressure scale height, and $c_v$ and $c_p$ are the heat capacities per unit mass at constant volume and constant pressure.
The viscous stress tensor
$\mathcal{D}_{ij} = \rho \nu_{\rm t} [ \partial_i u_j' + \partial_j u_i' - \frac{2}{3} (\partial_k {u}'_k) \delta_{ij} ]$ 
accounts for wave attenuation, where $\delta_{ij}$ is the Kronecker delta.
The energy equation includes advection and thermal diffusion. 
In our model, the viscous and thermal diffusivities are those resulting from the turbulence and, therefore, were considered to be equal.  

The latitudinal entropy gradient is obtained by assuming that the differential rotation is the result of a thermal wind balance \citep{Miesch2006}:
\begin{equation}
    \frac{g}{c_p}\frac{\partial s}{\partial\theta} = r^{2}\sin{\theta}\frac{\partial (\Omega^2)}{\partial z},
\end{equation}
where $z = r \cos \theta$ is the coordinate along the rotation axis. 

Boundary conditions need to be applied at $\theta = 0$ and $\pi$. Since we only considered modes with $m \neq 0$, we imposed $\bu' = 0$ and $\rho' = p' = s' = 0$. 
{The numerical domain is bounded above by the photosphere and below by the radiative interior, both of which are strongly stably stratified and where radial flows are difficult to drive because of the strong, restoring buoyancy force. }
{Therefore, we used an impenetrable and stress-free boundary condition at both radial boundaries. }

We looked for solutions to the above problem where each physical quantity is proportional to $\exp(\iim m \phi - \iim \sigma t)$, where $\sigma$ is the complex mode angular frequency and $m$ is the integer longitudinal wavenumber.
We discretized the spatial derivatives with second-order central differences with $16$~radial and $72$~latitudinal grid points. 
The above equations were combined in matrix form into a complex eigenvalue problem, which was solved using the LAPACK routine. We focused on the low-frequency solutions.
We refer to the modes of oscillation  obtained in this problem as the "modes of the 2D model".

\subsection{1D~eigenvalue solver}
\label{app.1dsolver}

We also considered fluid motions that are purely toroidal (horizontal),  restricted to a spherical surface of radius $r$. Keeping the density constant, the linearized momentum equations in a frame rotating at $\Omega_{\rm carr}$ are as follows:
\begin{align}
  & D_t  u'_\theta = 
  -\frac{1}{r}\frac{\partial}{\partial\theta} \left(\frac{p'}{\rho}\right) + 2 \Omega \cos\theta\ u'_\phi    + \nu_{\rm t}\  \Delta u'_\theta,  \\
  & D_t  u'_\phi = - \frac{1}{r\sin \theta}\frac{\partial}{\partial \phi} \left(\frac{p'}{\rho}\right) - 2 \Omega \cos\theta\ u'_\theta -  \sin\theta\ u'_\theta \frac{\partial \Omega}{\partial \theta} 
  + \nu_{\rm t}\  \Delta u'_\phi,
\end{align}
where the material derivative $D_t$ is given above by Eq.~\eqref{eq.materialDer}, and $\Delta$ is the horizontal part of the Laplacian. 
For purely toroidal modes, we introduced the stream function $\Psi(\theta, \phi, t)$ such that
\begin{equation}
    \bu' = \bnabla \times \left[ \Psi(\theta,\phi,t)\ \bm{\hat{r}}  \right] = \frac{1}{r \sin\theta} \frac{\partial \Psi}{\partial \phi}\  \bm{\hat{\theta}} - \frac{1}{r}\frac{\partial \Psi}{\partial \theta}\ \bm{\hat{\phi}}.
\label{eq.u'}
\end{equation}
The two above equations can be combined to obtain 
\begin{equation}
D_t \Delta \Psi -  \frac{1}{r^2\sin\theta} \frac{\partial}{\partial \theta} \left( \frac{1}{\sin\theta} \frac{\partial}{\partial \theta} (\Omega \sin^2\theta) \right) \frac{\partial \Psi}{\partial \phi} 
= \nu_{\rm t}\ \Delta^2\Psi .
\end{equation}
We looked for solutions of the form
\begin{equation}
    \Psi(\theta,\phi,t) = \textrm{Re} \left[ \psi(\theta) \exp(\iim m\phi - \iim \sigma t) \right],
\label{eq.Psi}
\end{equation}
where $m$ is the longitudinal wavenumber and $\sigma$ is the (complex) angular frequency. The equation for $\psi$ is of fourth-order and requires four boundary conditions. The condition that the flow vanishes at the poles implies
\begin{equation}
\psi = \frac{\mathrm{d} \psi}{\mathrm{d}\theta} = 0 \;
\textrm{ at\ $\theta = 0$ and $\pi$}.
\label{eq.Boundary}
\end{equation}
In order to discretize the problem, we projected $\psi$ onto a basis of associated Legendre polynomials. For the numerical value of the eddy viscosity at the surface, we used the value  $\nu_{\rm t} = 500$~km$^2$s$^{-1}$  \citep{Gizon2020-beta}, unless  otherwise specified.  The resulting eigenvalue problem was solved for each $m$ using the eigenvalue solver \verb1scipy.linalg.eig1.
We refer to the modes of oscillation  obtained in this problem as the "modes of the 1D model".

\section{Supplementary figures}

\begin{figure}[hb]
\centering
\includegraphics[width=0.7\linewidth]{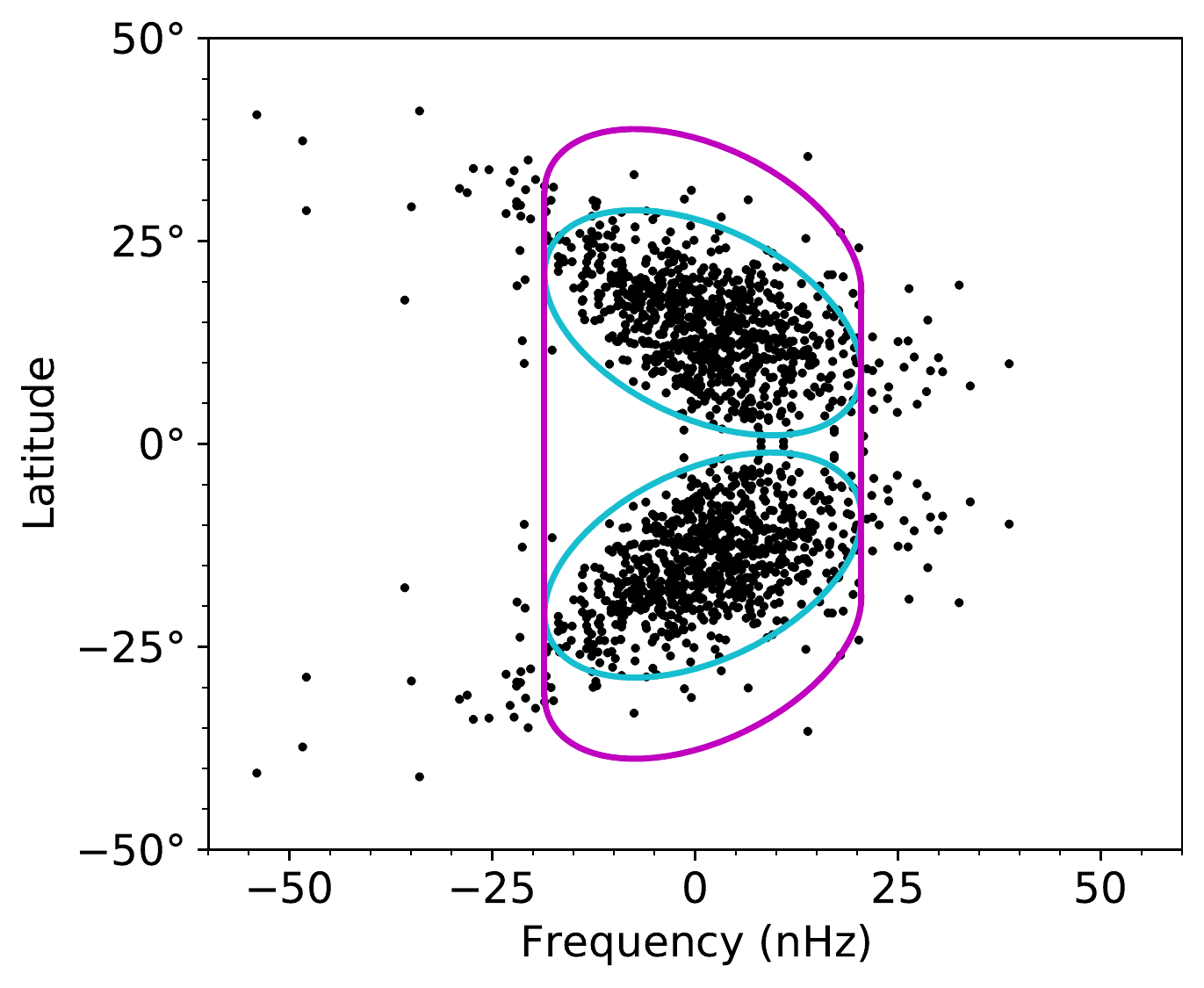}
\caption{Rotational frequencies of solar active regions versus latitude,   measured in the Carrington frame {\citep[from May 2010 to December 2016,][]{Kutsenko2021}}. The data (black dots) have been symmetrized in latitude. The cyan ellipses contain $90$\% of the active regions. The ellipses are extended to higher latitudes by $10^\circ$ and down to the equator to include flows around active regions. The resulting region in frequency--latitude space is given by the purple contour, which we denote via the equation $\omega = \Omega_{\rm AR}(\theta) - \Omega_{\rm Carr}$.
{We note that the HMI data used in the main text cover a longer observation period (from May 2010 to September 2020); however, the purple contour is not significantly affected by the very few active regions from the nearly quiet period 2017--2020 (about 5\% of all cycle 24 active regions).}
\label{fig.ar_freq_range}
}
\end{figure}

\begin{figure*}
\centering
\includegraphics[width=\linewidth]{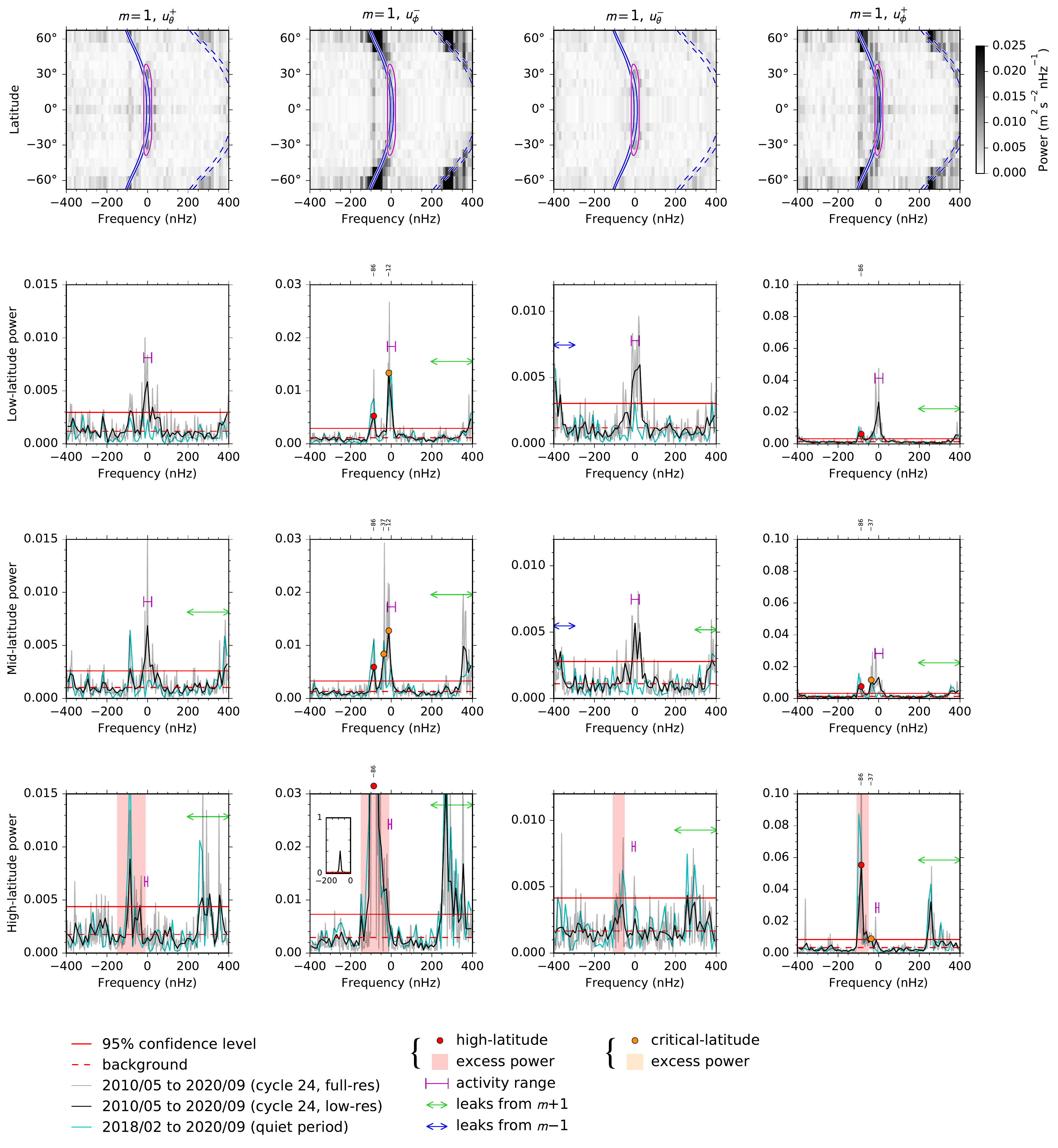}
\caption{Power spectra for $m=1$.
({\bf Top row}) Power for the four components $u_\theta^{+}$, $u_\phi^{-}$, $u_\theta^{-}$, and $u_\phi^{+}$. 
The purple contours delineate the regions where inflows into active regions produce excess power (see Fig.~\ref{fig.ar_freq_range}). 
The two blue curves show $m(\Omega - \Omega_{\rm Carr})/2\pi$ at the surface and at $r = 0.95 R_\odot$. 
({\bf Second row}) Power spectral density averaged over $0^\circ$\,--\,$30^\circ$.
 The gray curves show the power spectra at full resolution ($3.06$~nHz), and the black curves show them at a quarter of the resolution. 
The $95$\% confidence levels are shown by the red horizontal lines. The cyan curves are for the quiet-Sun period only ($2.6$~years from 4~February 2018 to 6~September 2020).  
({\bf Third row})
Power spectral density averaged over $15^\circ$\,--\,$45^\circ$.
({\bf Fourth row})
Power spectral density averaged over $37.5^\circ$\,--\,$67.5^\circ$.
In the three lower rows,
the dots and the shaded areas (see legend) indicate the significant peaks and the excess power ranges not related to magnetic activity, given in Table~\ref{full-table}.
\label{fig.overview.1}
}
\end{figure*}

\begin{figure*}
\centering
\includegraphics[width=\linewidth]{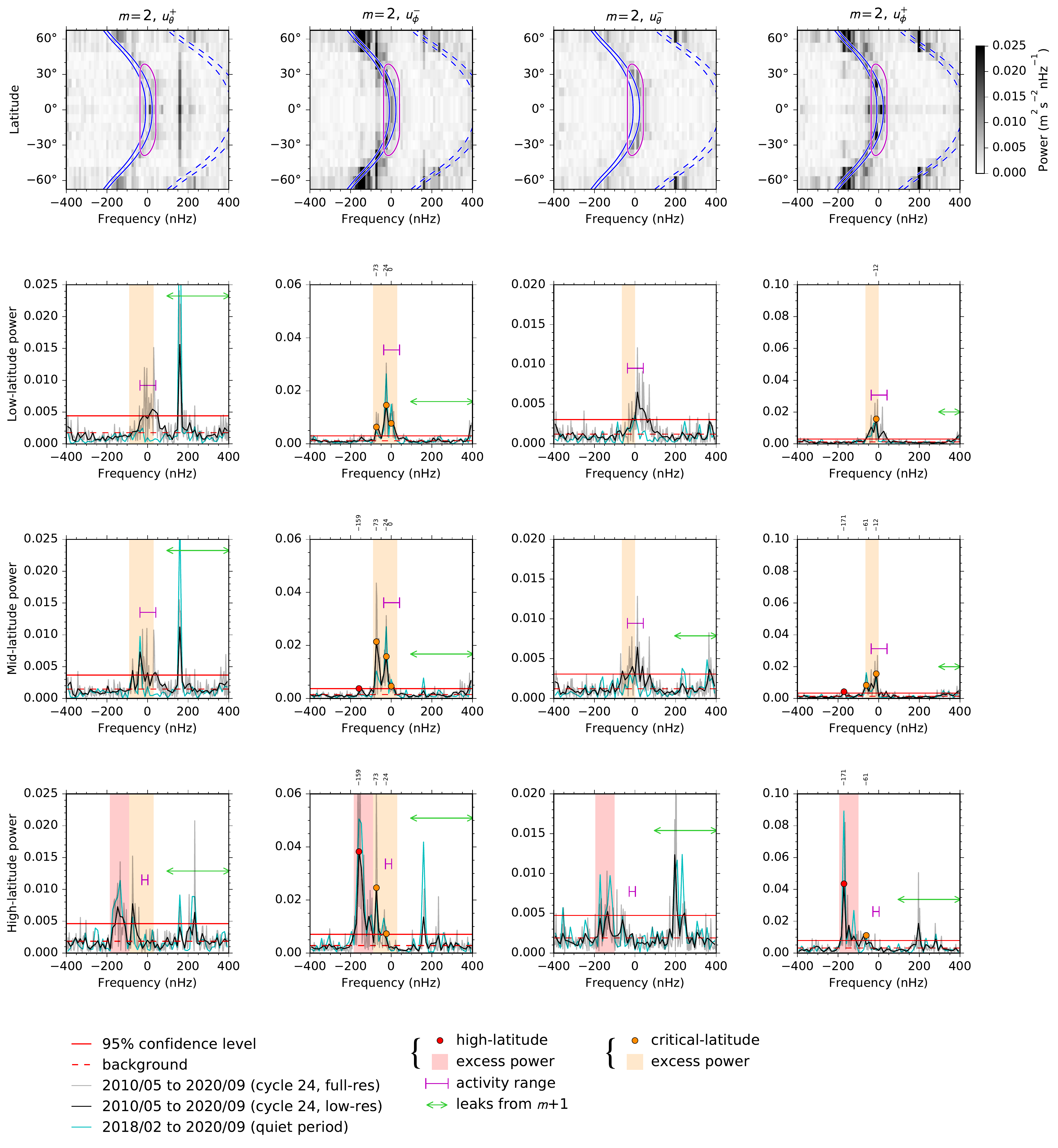}
\caption{Same as Fig.~\ref{fig.overview.1}, but for $m = 2$.
\label{fig.overview.2}
}
\end{figure*}

\begin{figure*}
\centering
\includegraphics[width=\linewidth]{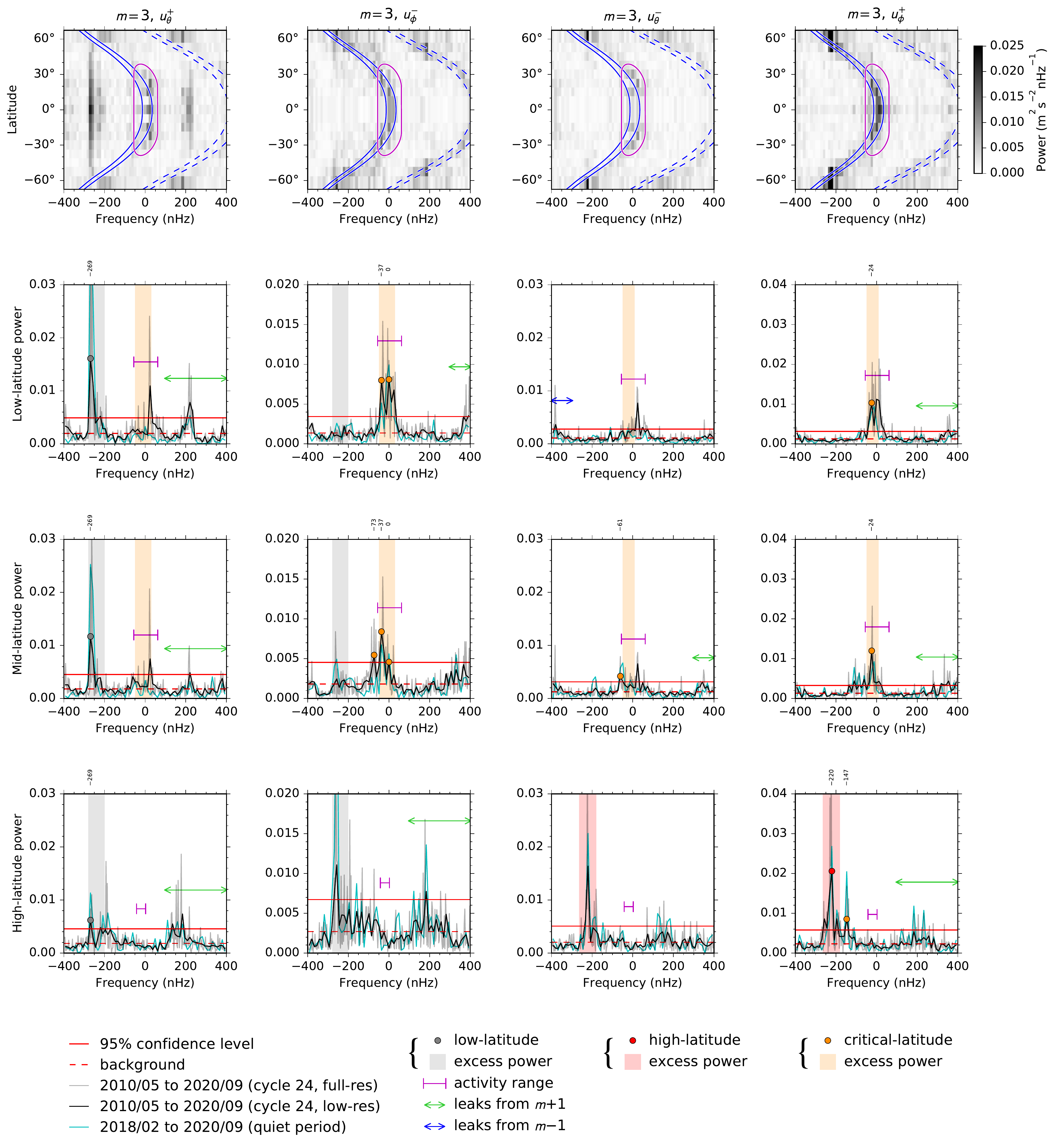}
\caption{Same as Fig.~\ref{fig.overview.1}, but for $m = 3$.
\label{fig.overview.3}
}
\end{figure*}

\begin{figure*}
\centering
\includegraphics[width=\linewidth]{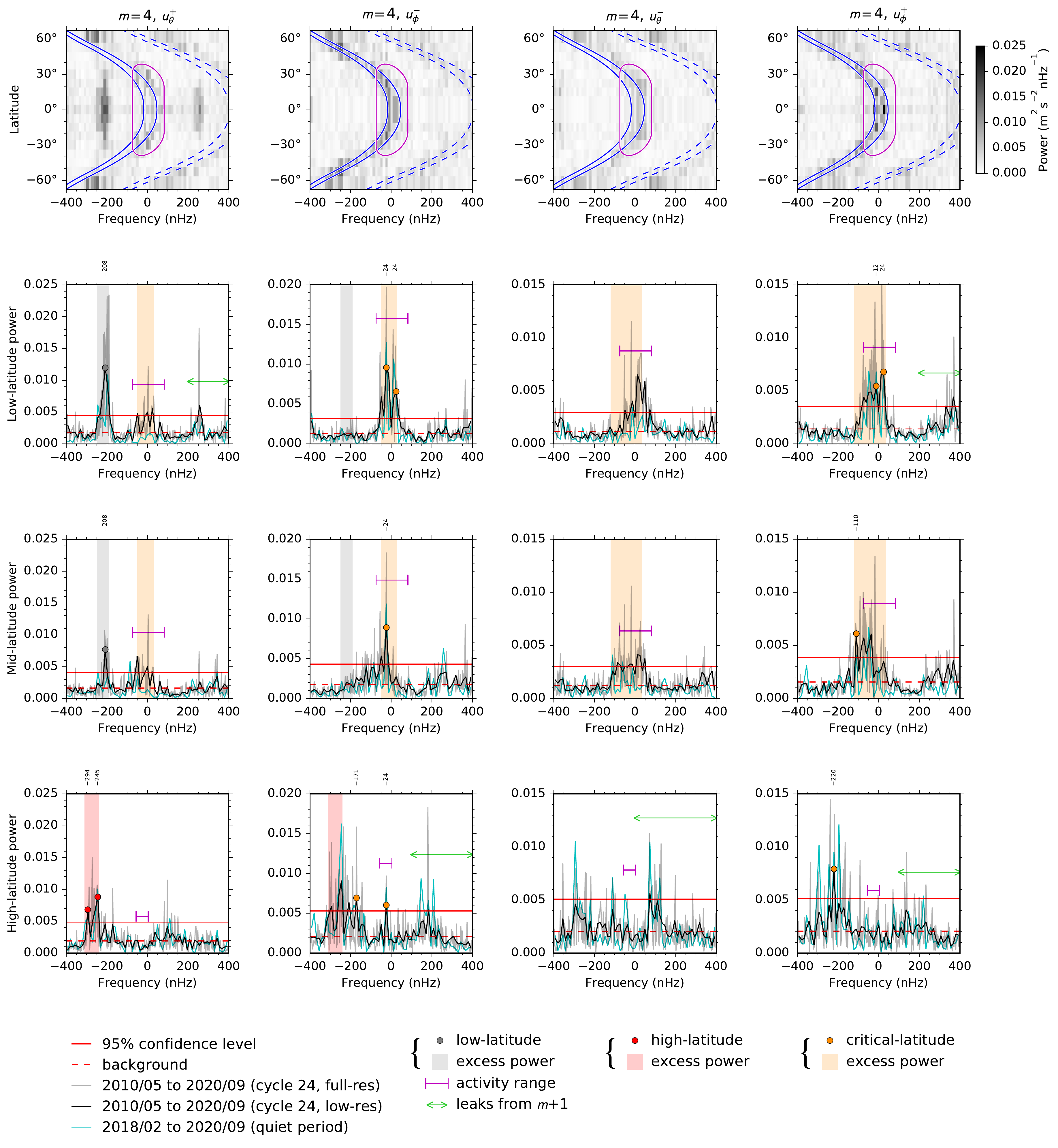}
\caption{Same as Fig.~\ref{fig.overview.1}, but for $m = 4$.
\label{fig.overview.4}
}
\end{figure*}

\begin{figure*}
\centering
\includegraphics[width=\linewidth]{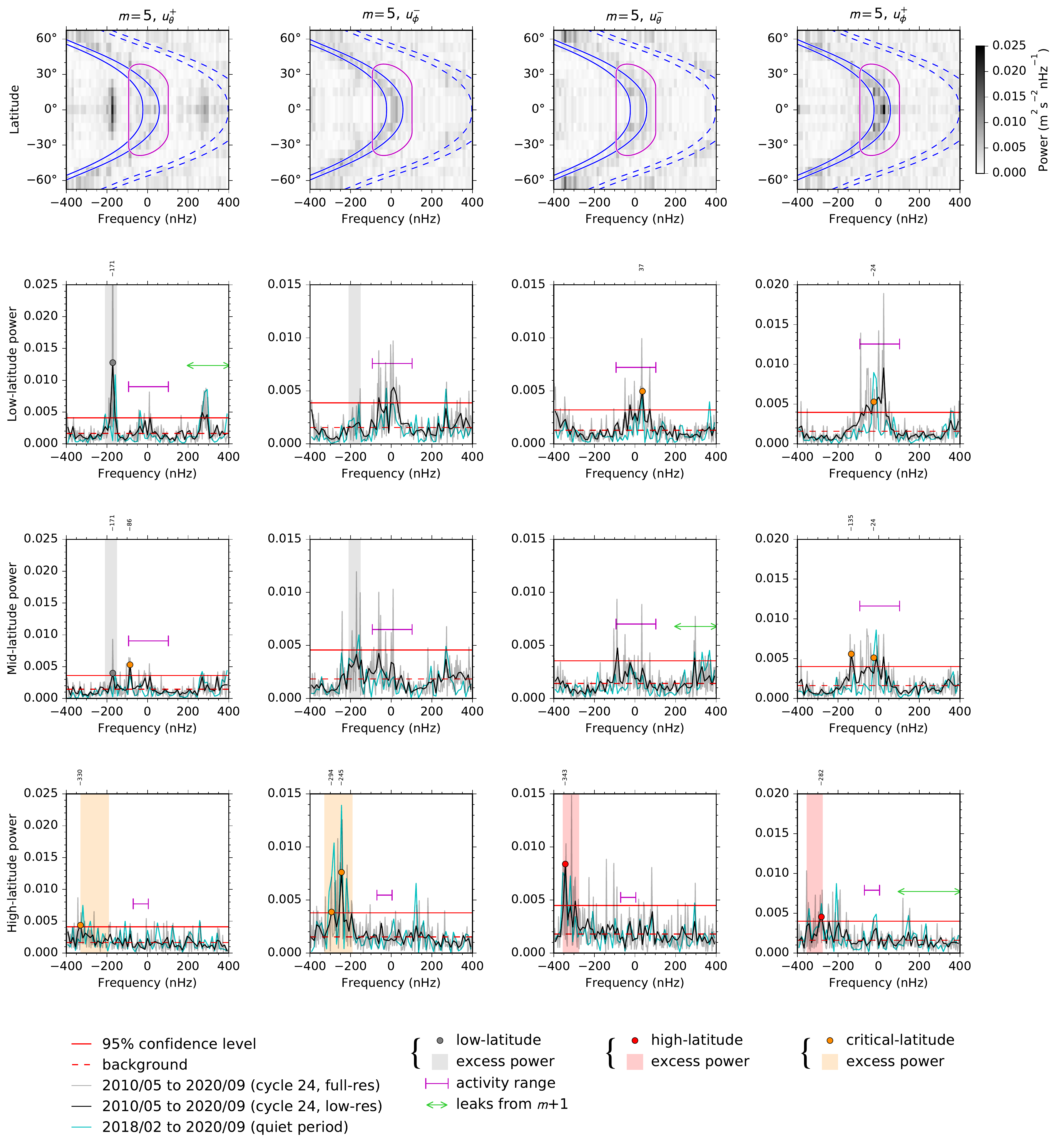}
\caption{Same as Fig.~\ref{fig.overview.1}, but for $m = 5$.
\label{fig.overview.5}
}
\end{figure*}

\begin{figure*}
\centering
\includegraphics[width=\linewidth]{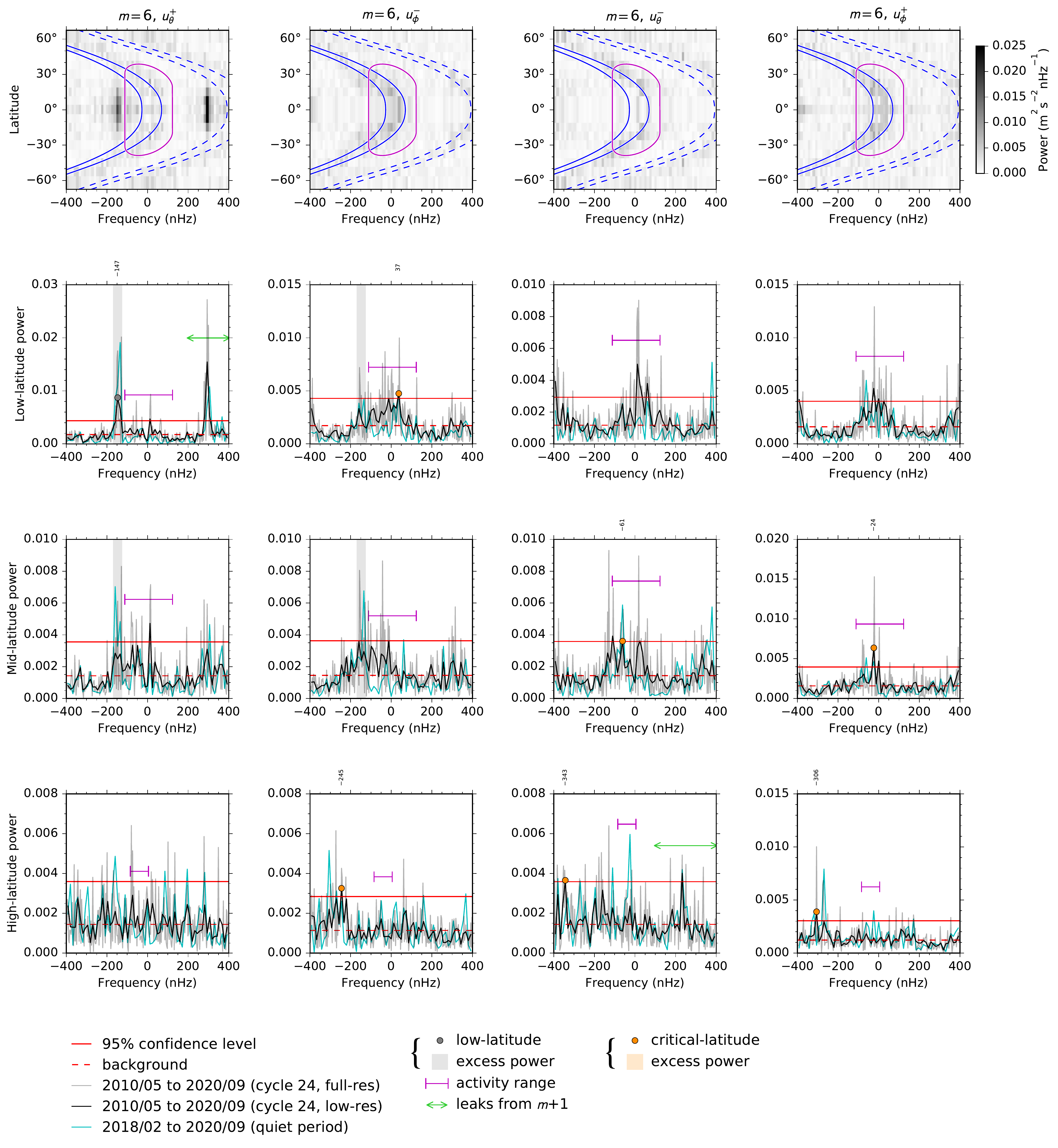}
\caption{Same as Fig.~\ref{fig.overview.1}, but for $m = 6$.
\label{fig.overview.6}
}
\end{figure*}

\begin{figure*}
\centering
\includegraphics[width=\linewidth]{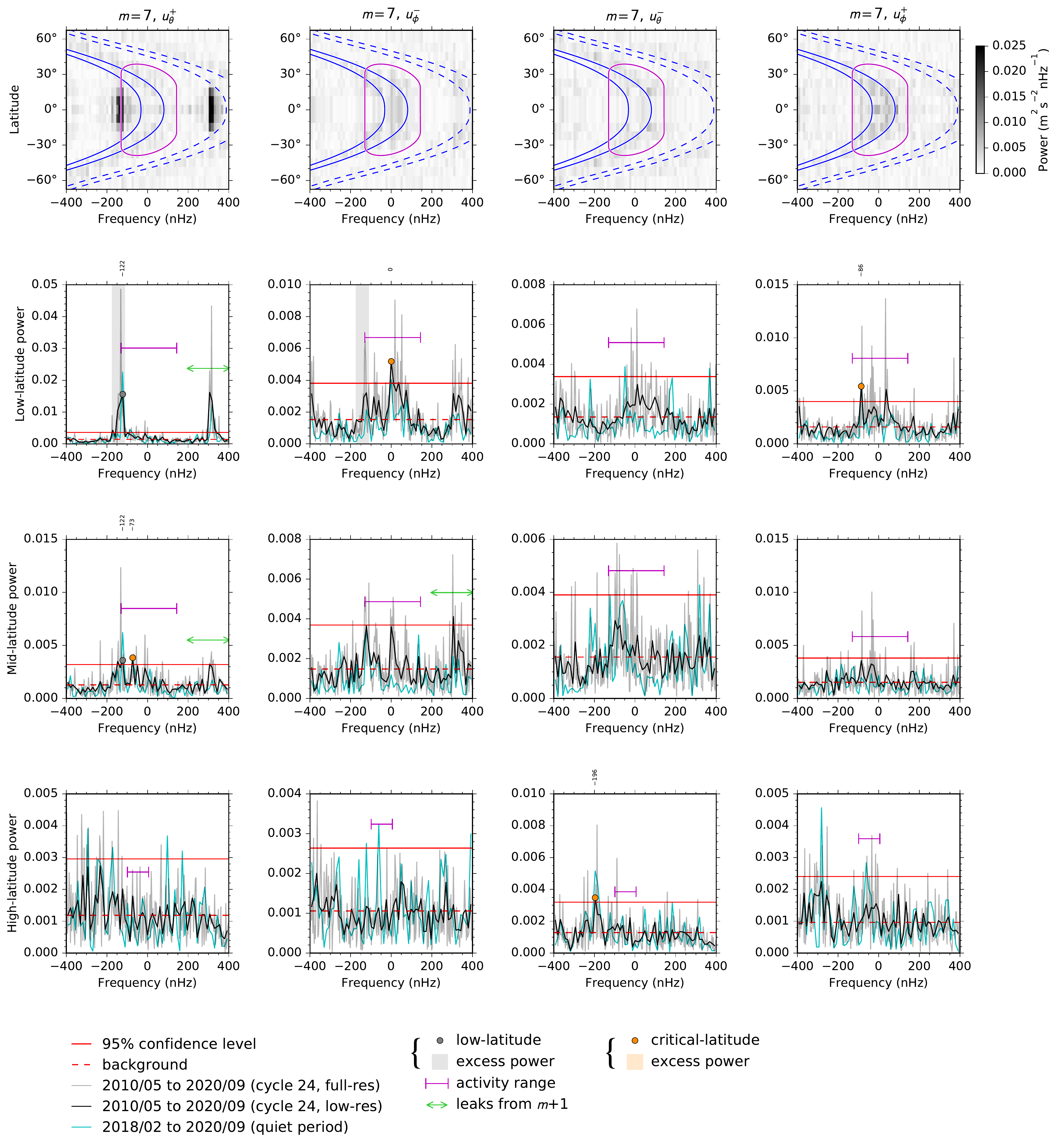}
\caption{Same as Fig.~\ref{fig.overview.1}, but for $m = 7$.
\label{fig.overview.7}
}
\end{figure*}

\begin{figure*}
\centering
\includegraphics[width=\linewidth]{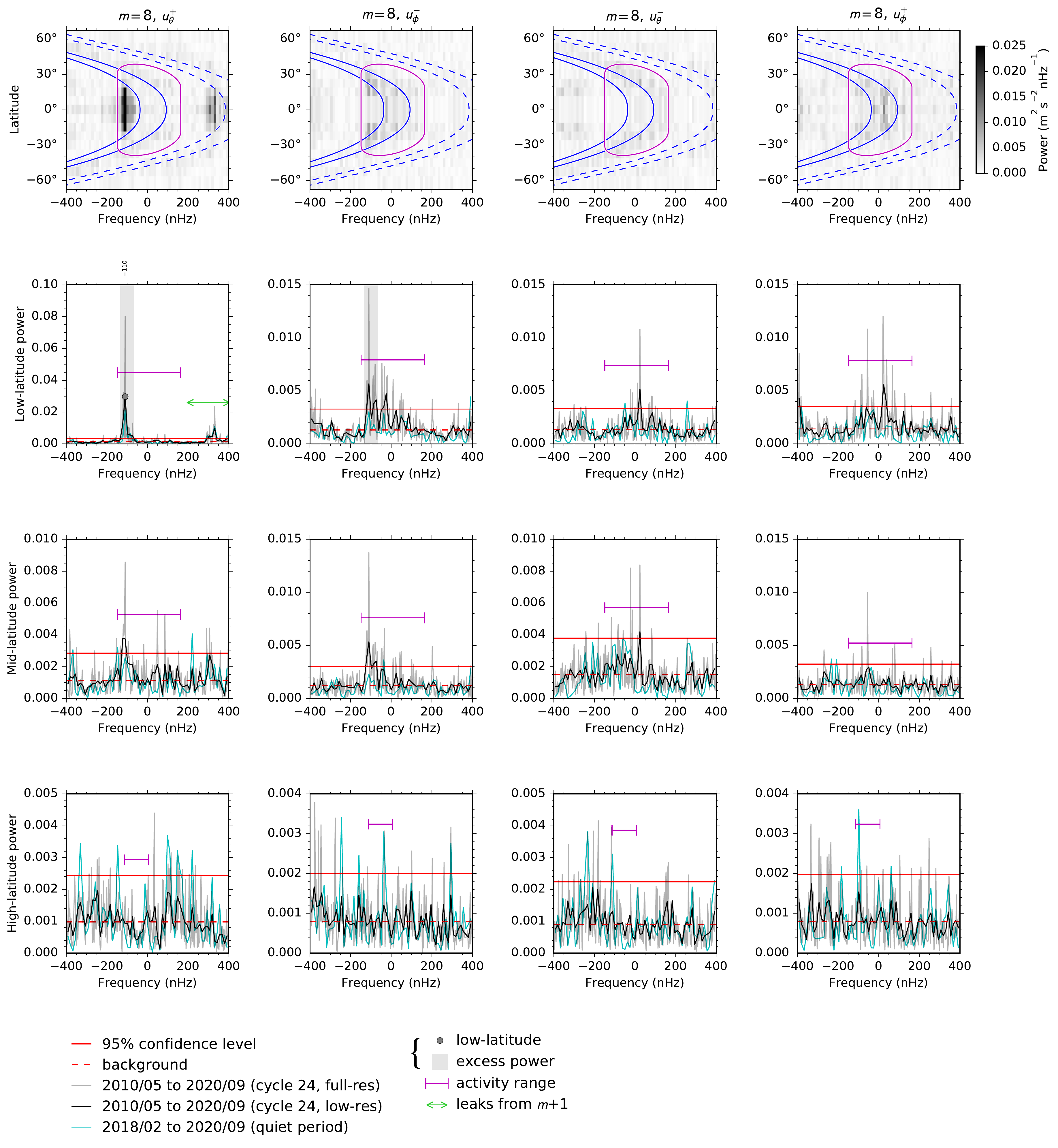}
\caption{Same as Fig.~\ref{fig.overview.1}, but for $m = 8$.
\label{fig.overview.8}
}
\end{figure*}

\begin{figure*}
\centering
\includegraphics[width=\linewidth]{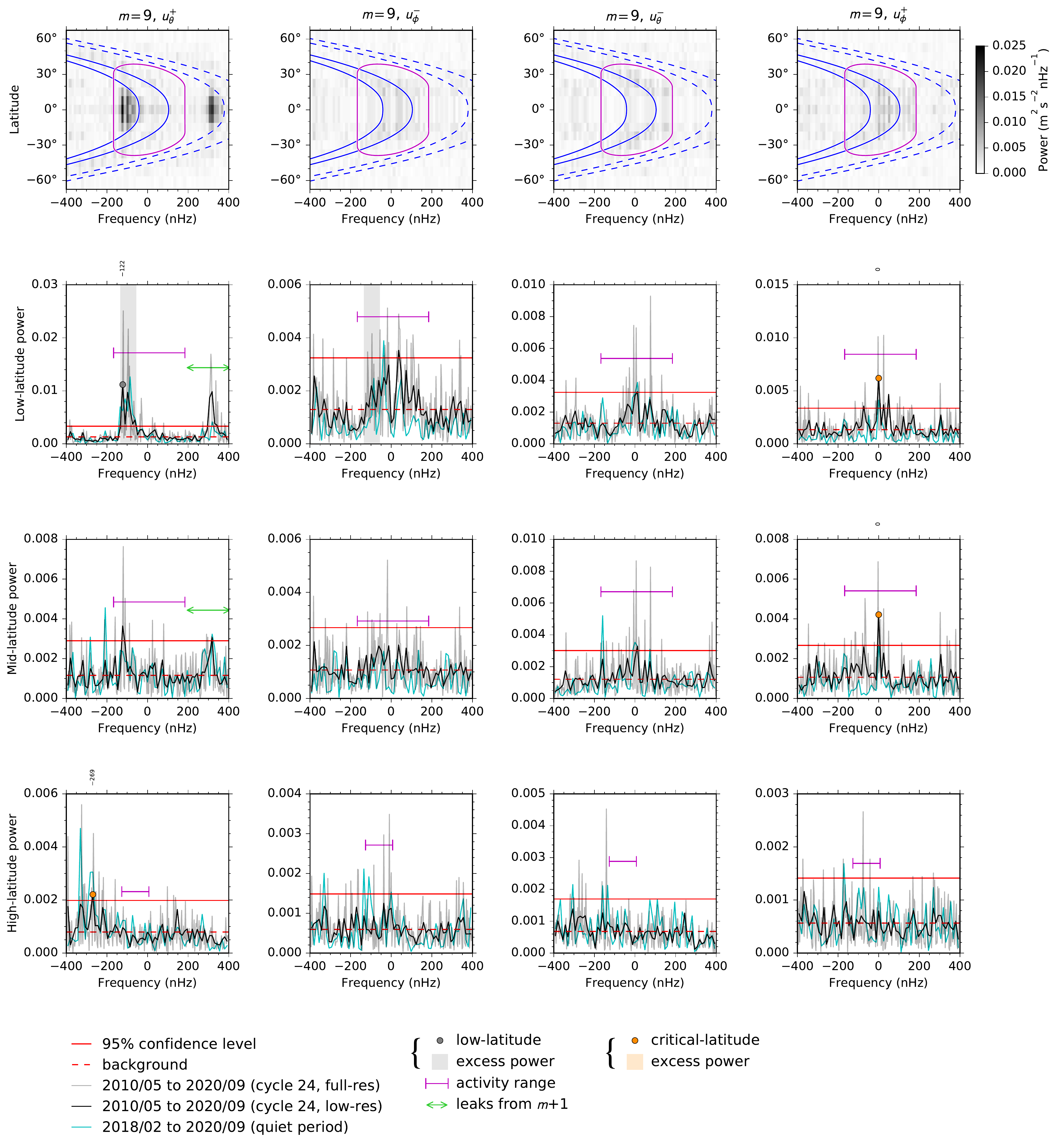}
\caption{Same as Fig.~\ref{fig.overview.1}, but for $m = 9$.
\label{fig.overview.9}
}
\end{figure*}

\begin{figure*}
\centering
\includegraphics[width=\linewidth]{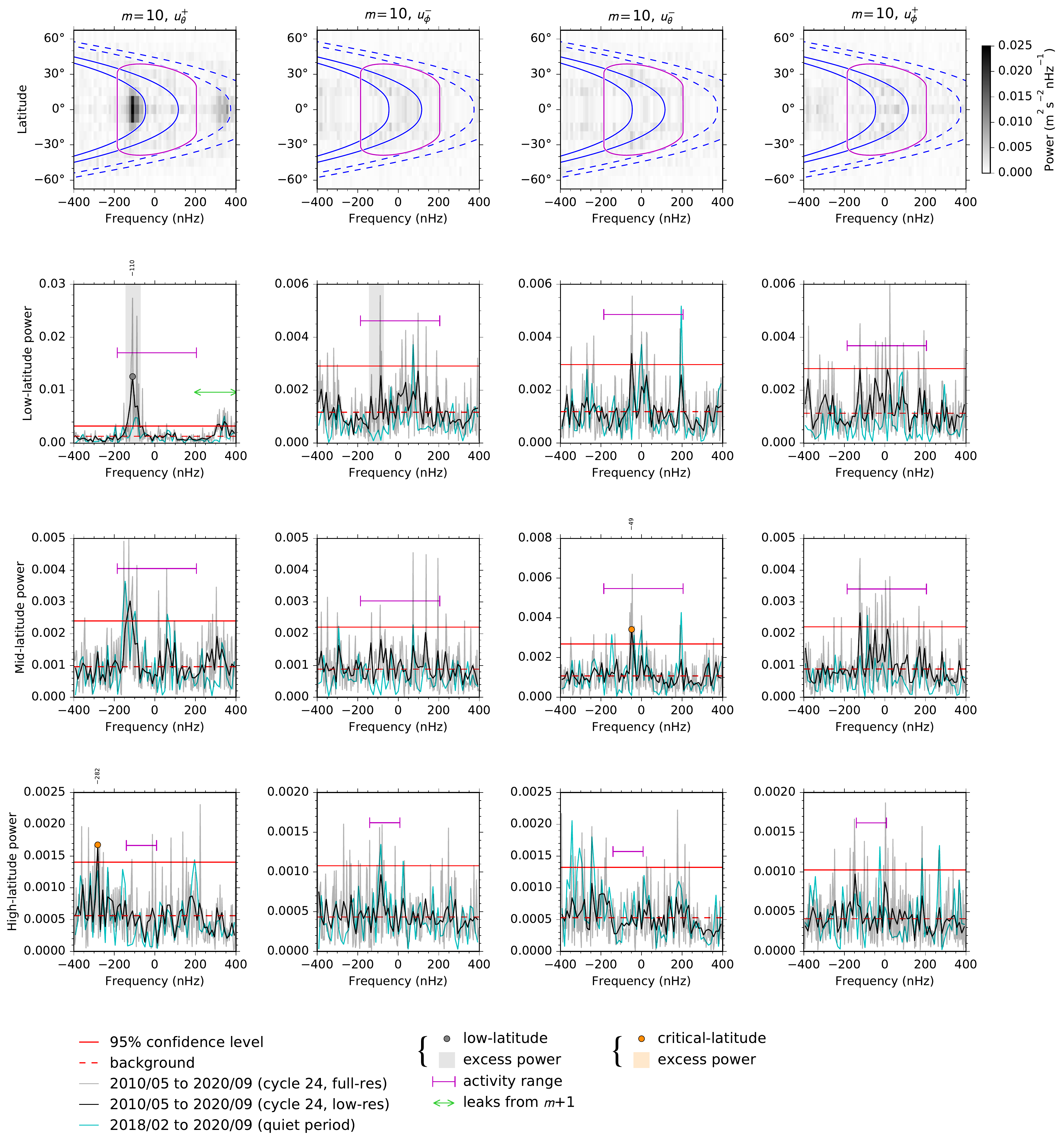}
\caption{Same as Fig.~\ref{fig.overview.1}, but for $m = 10$.
\label{fig.overview.10}
}
\end{figure*}

\begin{figure*} 
\centering
\includegraphics[width=0.8\linewidth]{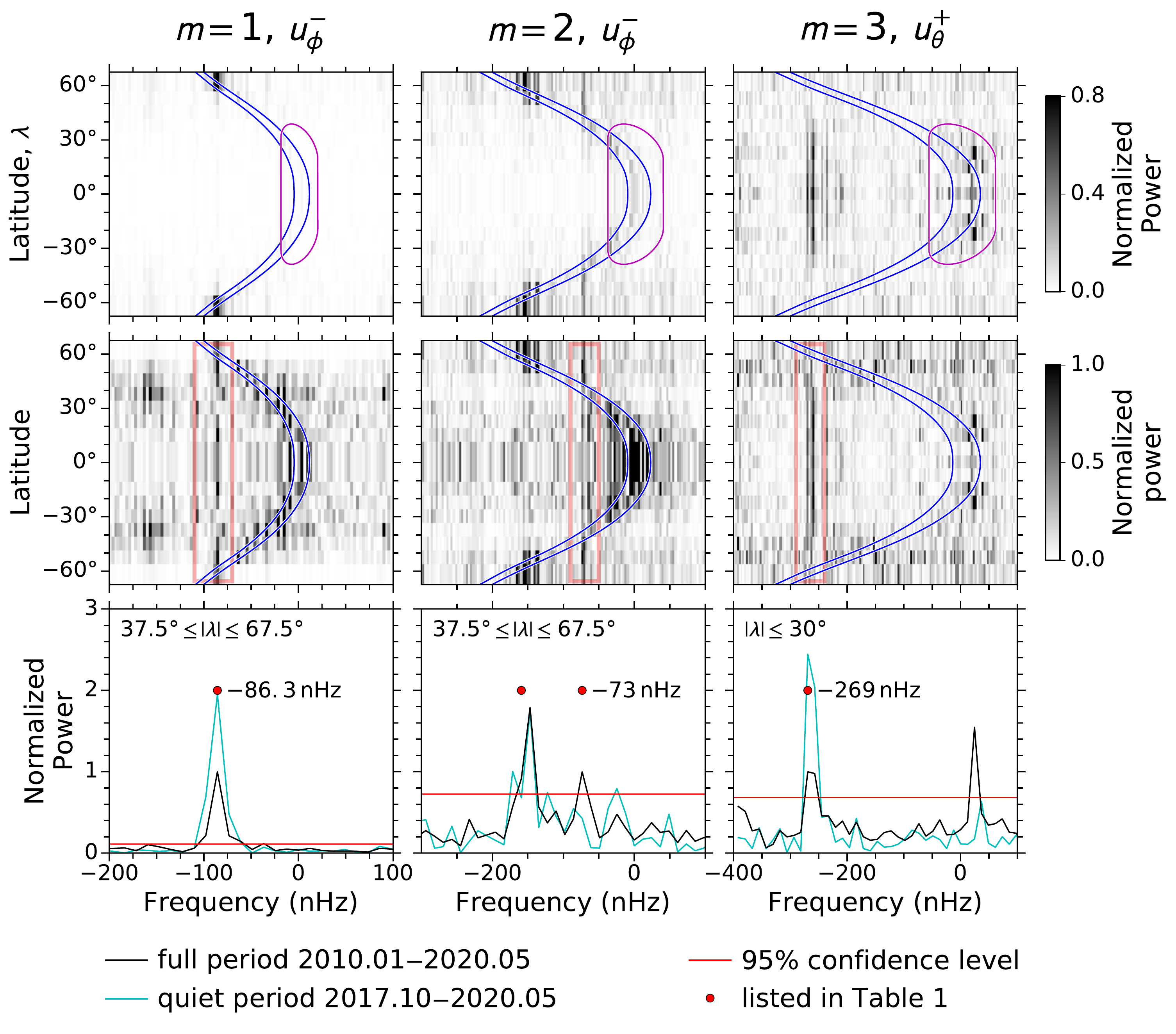}
\caption{
\label{fig.gong}
Same as Fig.~\ref{fig.observations}, but using GONG data.
The red dots mark the HMI frequencies for comparison.
}
\end{figure*}

\begin{figure*} 
\centering
\includegraphics[width=0.7\linewidth]{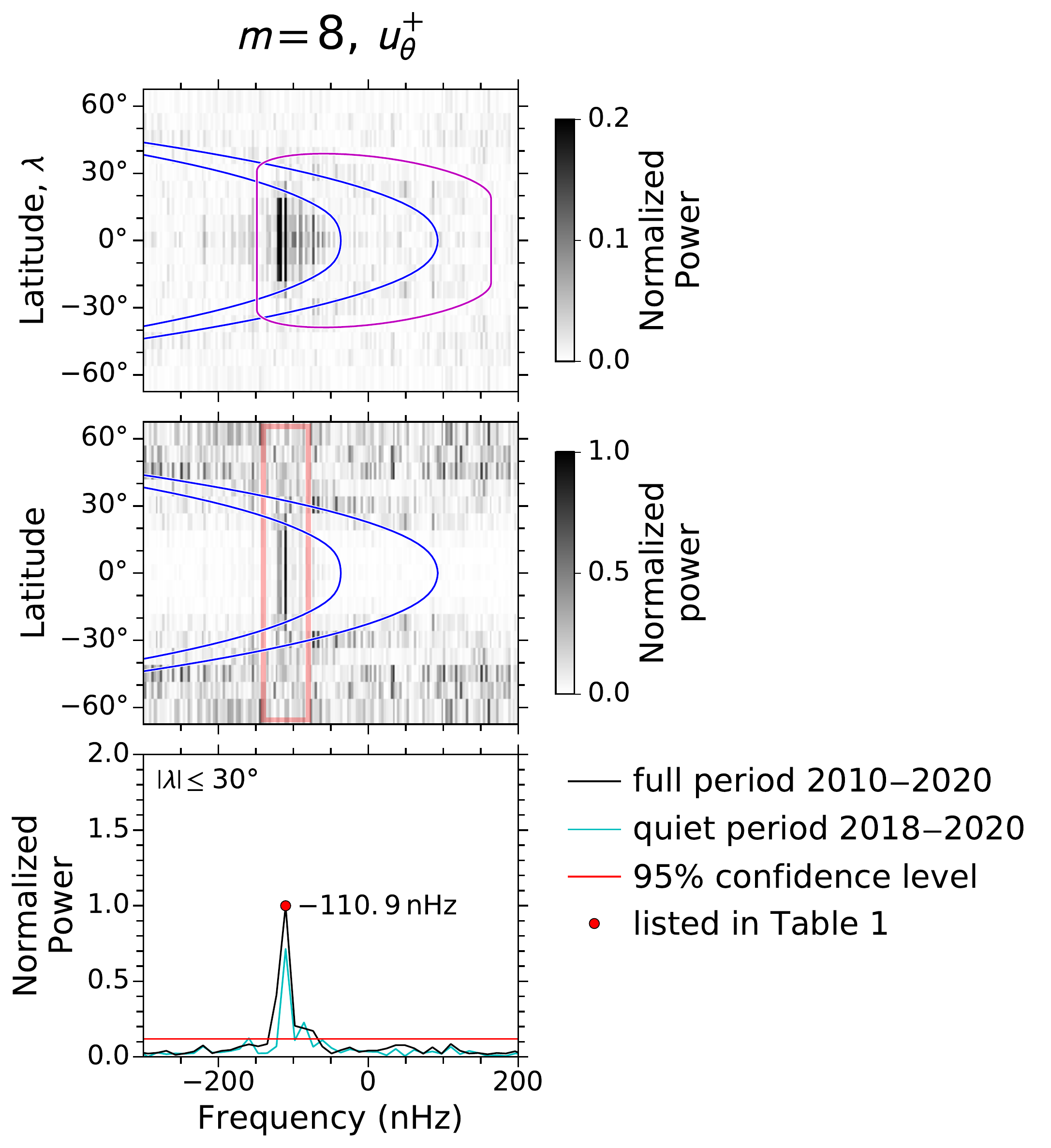}
\caption{
\label{fig.m8}
Same as Fig.~\ref{fig.observations}, but for the $m=8$ equatorial Rossby mode. In the top plot, it is important to 
notice the range of excess power at low latitudes, below the critical latitudes at the surface.
}
\end{figure*}

\clearpage
\begin{figure}
    \centering
    \includegraphics[width=0.7\linewidth]{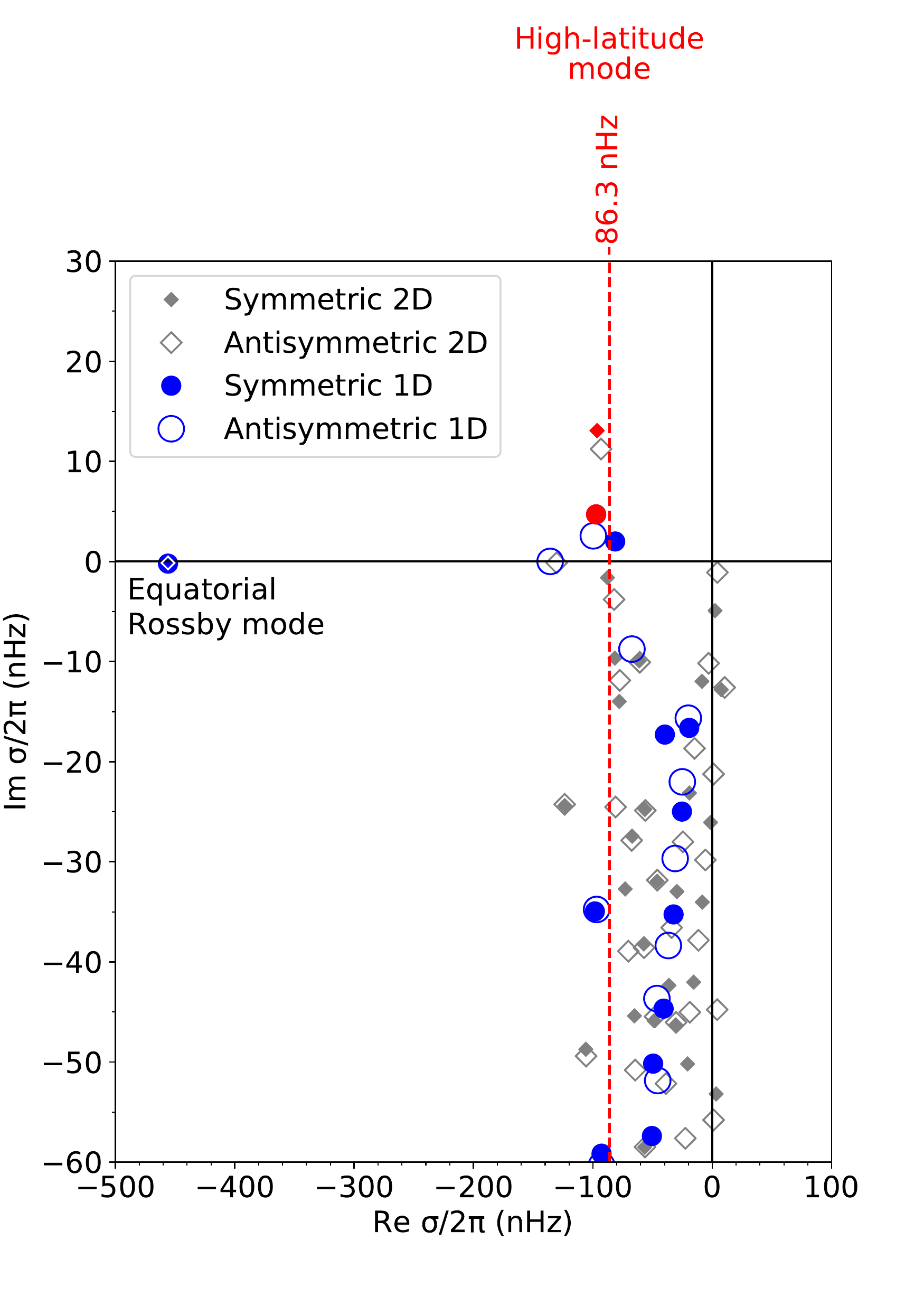}
    \caption{Eigenfrequencies in the complex plane for $m = 1$ from  the 2D~solver ($\delta=0$, $\nu_{\rm t}=250$~km$^2$\,s$^{-1}$) and the 1D~solver ($\nu_{\rm t}=250$~km$^2$\,s$^{-1}$). 
    Modes with positive imaginary frequencies are self-excited (unstable).
    The red vertical line shows the observed frequency of the high-latitude symmetric  mode at $-86.3$~nHz. The red symbols indicate the modes from the models, which have frequencies and surface eigenfunctions close to those observed.}
    \label{fig.eigenvalues-m1}
\end{figure}

\clearpage

\begin{figure*}
\centering
\includegraphics[width=0.985\linewidth]{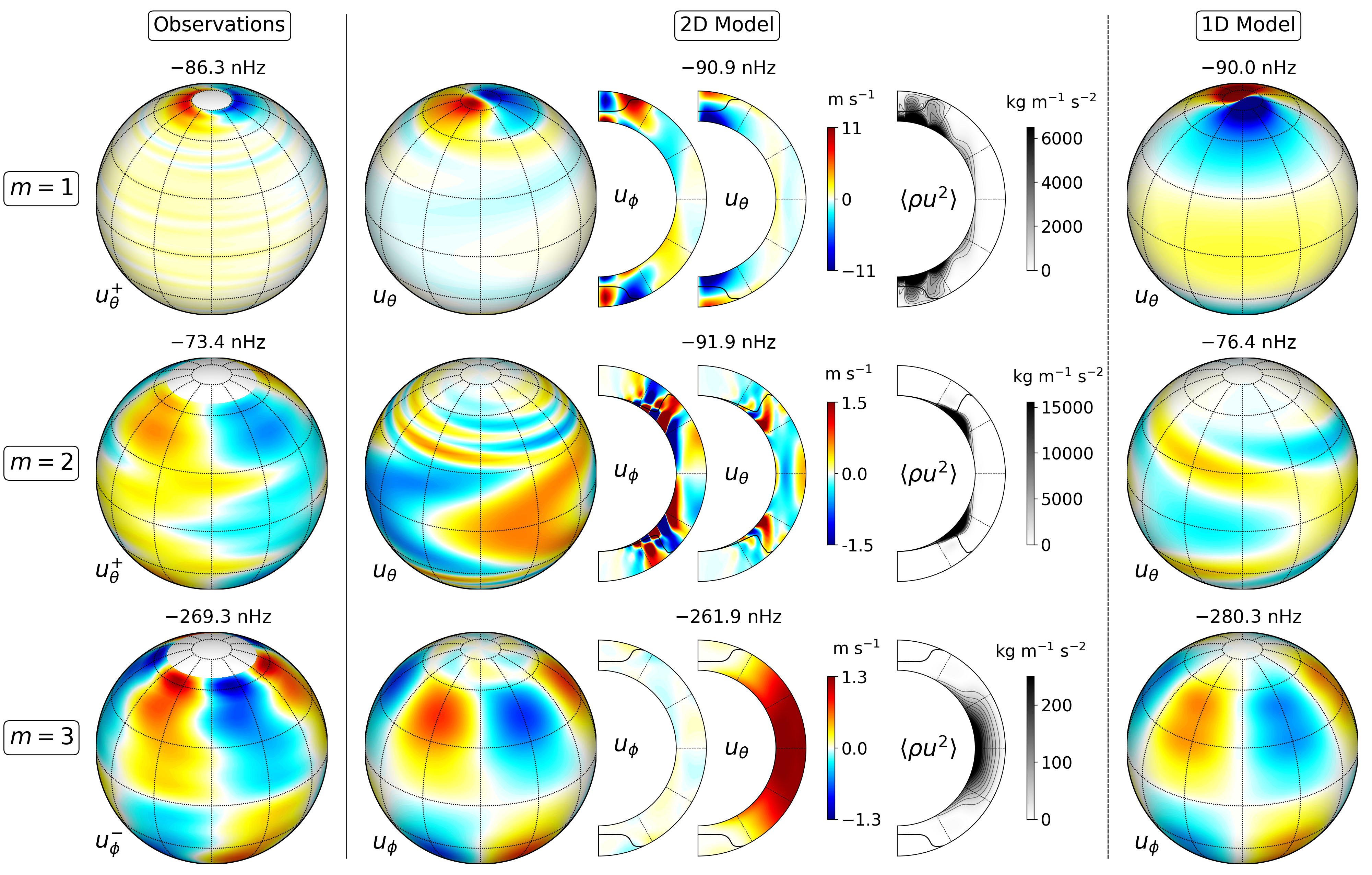}
\caption{Same as Fig.~\ref{fig.eigenfunctions}, but for the complementary velocity components.
\label{fig.eigenfunctions_complementary}}
\end{figure*}

\begin{figure*}
\centering
\includegraphics[width=0.99\linewidth]{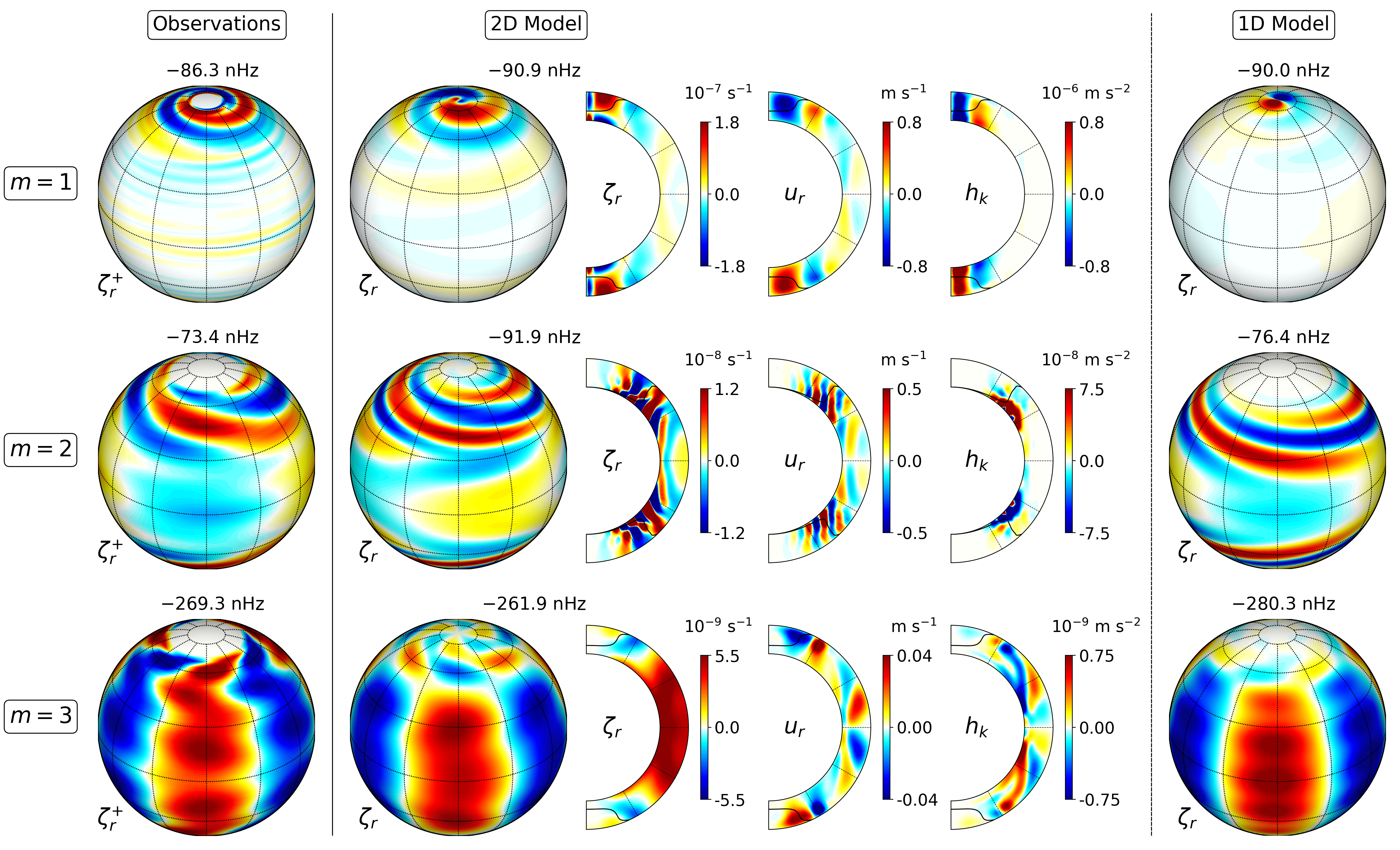}
\caption{
Observed and model radial vorticity for the selected modes of Fig.~\ref{fig.eigenfunctions}.
The first, second, and rightmost columns show the radial vorticity $\zeta_{r} = (\bnabla\times \bm{u})_{r}$ for the observations, the 2D~model, and the 1D~model, respectively.  The remaining columns in the middle show meridional cuts of $\zeta_{r}$, radial velocity $u_{r}$, and the kinetic helicity $h_{\rm k} = \langle \bm{u} \cdot \bm{\zeta} \rangle$ for the 2D~model.
\label{fig.vorticity}}
\end{figure*}

\begin{figure*}
\centering
\includegraphics[width=0.96\linewidth]{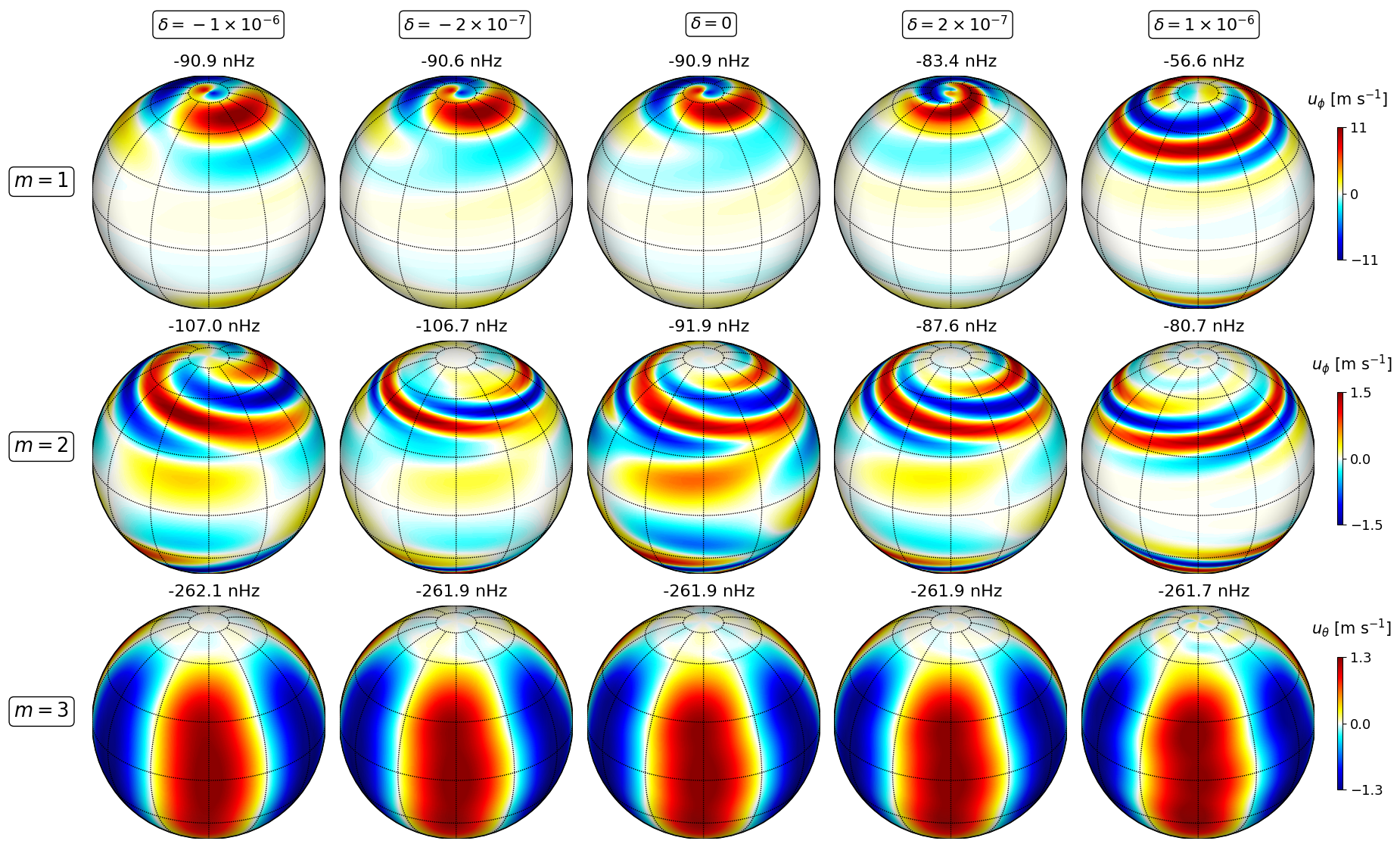}
\caption{Parameter study (2D~model) for different values of the superadiabaticity $\delta$, at fixed $\nu_{\rm t} = 100$~km$^2$ s$^{-1}$. The modes are those shown in Fig.~\ref{fig.observations}.
{The spiral patterns in $u_\phi$ of the $m=1$ high-latitude and  $m=2$ critical-latitude modes are sensitive to a small change in $\delta$. To obtain a pattern consistent with the observations,  $\delta < 2\times 10^{-7}$ is implied. 
The case $\delta=10^{-6}$ is excluded by both the eigenfunctions and the eigenfrequencies.
The $m=3$ equatorial Rossby mode is  almost independent of $\delta$ because it is nearly purely horizontal  (quasi-toroidal).}
\label{fig.changedelta}}
\end{figure*}

\begin{figure*}
\centering
\includegraphics[width=\linewidth]{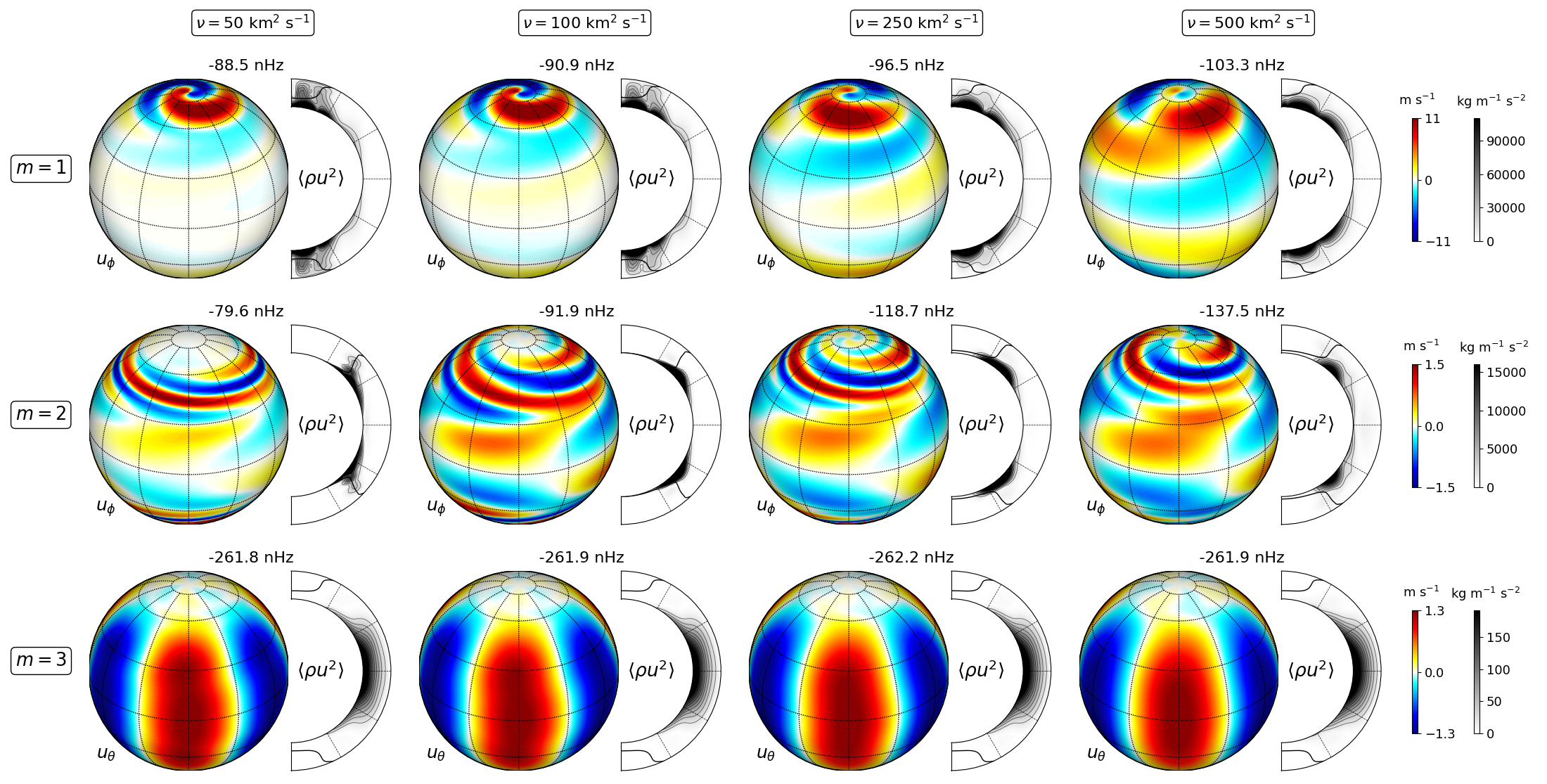}
\caption{
Parameter study (2D~model) for different values of the turbulent viscosity $\nu_{\rm t}$, for a convection zone that is adiabatically stratified ($\delta = 0$). 
The modes are those shown in Fig.~\ref{fig.observations}.
The frequencies of the $m=1$ high-latitude mode and the $m=2$ critical-latitude modes are sensitive to the choice of $\nu_{\rm t}$. The smaller values of $\nu_{\rm t}$ ($\leq 100$~km$^2$~s$^{-1}$) give a better agreement with the observed frequencies   (respectively $-86.3$ nHz and $-73.4$ nHz). The $m=3$ equatorial Rossby modes is essentially insensitive to $\nu_{\rm t}$.
\label{fig.changenu}}
\end{figure*}

\end{appendix}

\end{document}